\documentclass[reprint,twocolumn,superscriptaddress,showpacs,nofootinbib,notitlepage,showkeys,amsmath,amssymb,floatfix,preprintnumbers]{revtex4-1}
\usepackage[utf8]{inputenc}
\usepackage{graphicx}
\usepackage{latexsym,amsmath,amssymb,lmodern,float,url}
\usepackage{natbib}
\usepackage{color}
\usepackage{microtype}
\usepackage{slashed}
\usepackage{multirow}
\usepackage{bm}
\usepackage{array}

\usepackage[colorlinks=true,backref=false, linktocpage=true,
citecolor=blue,urlcolor=blue,linkcolor=blue,pdfpagemode=UseOutlines]{hyperref}

\hypersetup{%
  bookmarksnumbered=true,
  pdftitle = {},
  pdfsubject = {},
  pdfauthor = {},
  pdfkeywords = {}
}

\newcommand{\rjp}{R(J/\psi)}
\newcommand{\jp}{J/\psi}
\newcommand{\bc}{B_c^+}

\def\h3{{\textstyle{\frac 3 2}}}

\begin{document}
 
\title{Tests of the Standard Model in $B \to D\ell \nu_\ell$,
$B \to D^* \ell \nu_\ell$ and $B_c \to J/\psi \, \ell \nu_\ell$}

\author{Thomas D. Cohen}
\email{cohen@physics.umd.edu}
\author{Henry Lamm}
\email{hlamm@umd.edu}
\affiliation{Department of Physics, University of Maryland, College
Park, Maryland 20742-4111, USA}

\author{Richard F. Lebed}
\email{richard.lebed@asu.edu}
\affiliation{Department of Physics, Arizona State University, Tempe,
Arizona 85287-1504, USA}

\date{June, 2018}

\begin{abstract}
A number of recent experimental measurements suggest the possibility
of a breakdown of lepton ($\ell$) universality in exclusive $b \to c
\ell \nu_\ell$ semileptonic meson decays.  We analyze the full
differential decay rates for several such processes, and show how to
extract combinations of the underlying helicity amplitudes that are
completely independent of $m_\ell$.  Ratios of these combinations for
different $\ell$ (as well as some combinations for a single value of
$\ell$) therefore equal unity in the Standard Model and provide
stringent tests of lepton universality.  Furthermore, the extractions
assume the form of weighted integrals over the differential decay
rates and therefore are useful even in situations where data in some
regions of allowed phase space may be sparse.
\end{abstract}



\maketitle

\section{Introduction}
\label{sec:Intro}

The Standard Model (SM) has historically worked extremely well, but
many compelling reasons lead one to expect the existence of
beyond-Standard Model (BSM) physics.  Besides gravity, neutrino
oscillation is the only confirmed BSM physics, and certainly provides
significant information.  But it is important to seek out additional
regimes in which the SM fails, both for its own discovery potential
and to test our understanding of processes that have traditionally
been well understood in the SM.

One of the most prominent and intriguing experimental tensions with
the SM at present appears in the semileptonic decays of $B$ mesons and
of $B_c$ mesons, {\it i.e.}, $B \to D \ell \nu_\ell$, $B \to D^* \ell
\nu_\ell$, and $B_c \to J/\psi \, \ell \nu_\ell$, where $\ell$ is a
generic charged lepton.  The tension arises when comparing the ratio
$R(H)$ of total $\ell \! = \! \tau$ to total $\ell \! = \! \mu,e$
decay rates, where $H$ is the daughter hadron.  The HFLAV
averages~\cite{Amhis:2016xyh} of the experimental values for the $B$
decays are
$R(D)=0.407(39)(24)$~\cite{Lees:2012xj,Lees:2013uzd,Huschle:2015rga}
and $R(D^*)=0.304(13)(7)$\cite{Lees:2012xj,Lees:2013uzd,
Huschle:2015rga,Sato:2016svk,Aaij:2015yra,Hirose:2016wfn,
Wormser:2017hsx}.  At present, only LHCb has measured the value of
$R(J/\Psi)=0.71(17)(18)$~\cite{Aaij:2017tyk}.  These values are
compared with results of SM calculations: The value
$R(D)=0.300(8)$~\cite{Aoki:2016frl} is an average of lattice QCD
results~\cite{Lattice:2015rga,Na:2015kha}, which can be combined with
measured form factors to reduce the uncertainty, leading to
$R(D)=0.299(3)$~\cite{Bigi:2016mdz}.  Using only the experimental form
factors from Belle~\cite{Dungel:2010uk}, $R(D^*)=0.252(3)$ was
computed in~\cite{Fajfer:2012vx}.  With the preliminary
$B_c^+\rightarrow J/\Psi$ lattice QCD results
of~\cite{Colquhoun:2016osw,*ALE}, a 95\% confidence level (CL) region
of $0.20\leq\rjp\leq0.39$ can be obtained~\cite{THRprep}.  The
discrepancies with the SM predictions are 2.3$\sigma$, 3.5$\sigma$,
and 1.3$\sigma$, respectively.  Moreover, the HFLAV combined analysis
of $R(D)$ and $R(D^*)$ yields a 4.1$\sigma$
discrepancy~\cite{Amhis:2016xyh}.  Recent $R(D^*)$ results from
LHCb~\cite{Wormser:2017hsx} and Belle~\cite{Hirose:2017dxl} suggest a
value more consistent with theory, but at present are unincorporated
into the global fit.

Of course, this tension could be due to statistical fluctuations
and/or some subtle systematic experimental bias.  If, however, these
results are early signals of BSM physics, then a natural explanation
could be a breakdown of {\em lepton universality}, {\it i.e.}, some
process by which the $\tau$ and $\nu_\tau$ couple to the decaying $B$
or $B_c$ meson differently than do a $\mu$ and $\nu_\mu$.
Accordingly, it is useful to construct more experimental tests of
lepton universality, beyond just $R(H)$.  The value of such tests lies
in their utility to isolate where the apparent violation of the SM
arises.

In principle, obtaining more sensitive tests is straightforward.
$B$-meson decays depend upon the 4-momentum and spin state of $\ell$
and the decay products of the final hadrons.  The process is thus
characterized by a differential decay rate expressed in terms of many
variables (angles, momentum transfers, {\it etc.}).  In the absence of
BSM physics, the entire differential decay rate is predicted by the
SM\@.  If these predictions are known with sufficient precision, a
direct comparison to the $\tau$ and $\mu$ rates from experimental data
serves as a test of the SM, allowing one to see precisely where the SM
breaks down.

There are, however, two major practical difficulties in implementing
such a scheme.  The first is the requirement of a full prediction from
the SM\@.  While to good approximation one can ignore higher-order
electroweak effects in semileptonic decays, a SM prediction requires
knowledge of several transition form factors of the $B$ ($B_c$) to the
$D^{(*)}$ ($J/\psi$).  These form factors involve strong interactions,
preventing perturbative calculations, but they are amendable to
lattice QCD\@.  At present, only the $B\rightarrow D$ form factors
have been computed with a complete treatment of
uncertainties~\cite{Lattice:2015rga, Na:2015kha}.  Partial results
exist for $B\rightarrow
D^*$~\cite{Bailey:2014tva,Harrison:2016gup,Aviles-Casco:2017nge,
Harrison:2017fmw,Bailey:2017xjk} and $\bc\rightarrow
\jp$~\cite{Colquhoun:2016osw,*ALE}, but do not cover the entire
allowed range of momentum transfer or have control of their
systematics.  Even with these limited results, combined constraints on
$R(H)$ can be made by application of dispersive relations and heavy
quark symmetries~\cite{Bigi:2017jbd,THRprep}.

While ignorance of the form factors yields a degree of uncertainty in
the prediction of $R(H)$, the estimates of these uncertainties have
relatively mild consequences for this ratio---provided the form-factor
determinations can be trusted.  The same cannot be said of the
differential decay rates, with all of their parametric dependences.
 
With sufficient data, one might hope to extract the form factors
directly and then check for self-consistency with the SM\@.  For
example, one could extract the form factors from the $\mu$ channel and
then use these to predict the differential decay rate for the $\tau$
channel.  A comparison of the predicted differential decay rate with
the experimental one would then probe the SM\@.  However, this
approach is difficult because it requires a considerable amount of
reliable data to implement.  To be successful, one would need to
extract the differential decay rate above experimental background with
reasonable accuracy over all allowed ranges of all kinematic
variables.

In this paper we propose a number of tests of the SM that are
particularly sensitive to lepton universality violations in $b \!
\rightarrow \! c$ semileptonic meson decays.  These tests
directly probe lepton universality, while having the virtue of being
form-factor independent.  Moreover, it is likely that some of the
proposed tests can be implemented with relatively sparse data.  The
basic method is to consider the ratio of the $\tau$ to $\mu$ channels
of particular weighted integrals of the differential decay rates.
These ratios equal unity in the SM (up to subleading electroweak
corrections), and their deviation from unity constitutes a measure of
the violation of lepton universality.  The robustness of these tests
lies in the choice of weight functions: Although the hadronic form
factors may be unknown, their momentum transfer ($q^2$) dependence is
identical for the $\tau$ and $\mu$ channels.

The tests probe universality for the following basic reason: In the SM
these decays are dominated by the decay of the $B$ ($B_c$) meson into
a $D^{(*)}$ ($J/\psi$) via the emission of a virtual $W$, which
subsequently decays into the charged lepton $\ell$ and neutrino
$\nu_\ell$.  The processes in which the final lepton is a $\tau$ or
$\mu$ are distinguished only by the kinematics associated with the
different $m_\ell$.  However, these kinematical differences lead to
different weightings of the various form factors, even at the same
value of $q^2$.  If instead, one takes special kinematically weighted
averages over the differential decay rates, then lepton universality
of the SM requires that these averages are equal.

In addition to testing for violations of lepton universality, we
construct other SM tests that do not require knowledge of the form
factors.  These tests are ratios of weighted
integrals of the differential decay rate, but can be performed using
a {\em single\/} type of lepton $\ell$.

This work is by no means the first attempt to overcome the
difficulties of extracting useful information from the full
differential decay rates.  Prior works~\cite{Colangelo:2018cnj,
Ivanov:2016qtw,Tran:2018kuv,Bhattacharya:2015ida,Bhattacharya:2016zcw,
Bhattacharya:2018kig, Jaiswal:2017rve} with different aims ({\it
e.g.}, to study the effect of form-factor parameterizations,
generalized BSM studies, and effects of the polarization of the $D^*$)
have tackled similar problems.  In particular, the use of helicity
amplitudes (which are particular linear combinations of form factors)
are employed in many of these works, as well as in the present paper.
Moreover, the ``trigonometric moments'' of Ref.~\cite{Ivanov:2016qtw}
are closely related but not identical to the weight integrals used
here.
 
This paper is organized as follows: Sec.~\ref{sec:BD} describes a set
of possible experimental tests of lepton universality and other
aspects of the SM for $B \to D \ell \nu_\ell$ and $B \to D^* \ell
\nu_\ell$.  The derivation of these tests depends upon the connection
of the differential decay rate to the helicity amplitudes, which are
described in detail in Sec.~\ref{sec:HA}\@.  Section~\ref{sec:Bc}
contains the tests for violations of lepton universality in $B_c \to
J/\psi \, \ell \nu_\ell$ and their derivation in terms of helicity
amplitudes.  Section~\ref{sec:Concl} contains closing remarks.

\section{Standard Model Tests in  $B \to D \ell \nu_\ell$ and
$B \to D^* \ell \nu_\ell$}
\label{sec:BD}

\begin{figure}[htb]
\begin{center} \includegraphics[width = \linewidth]{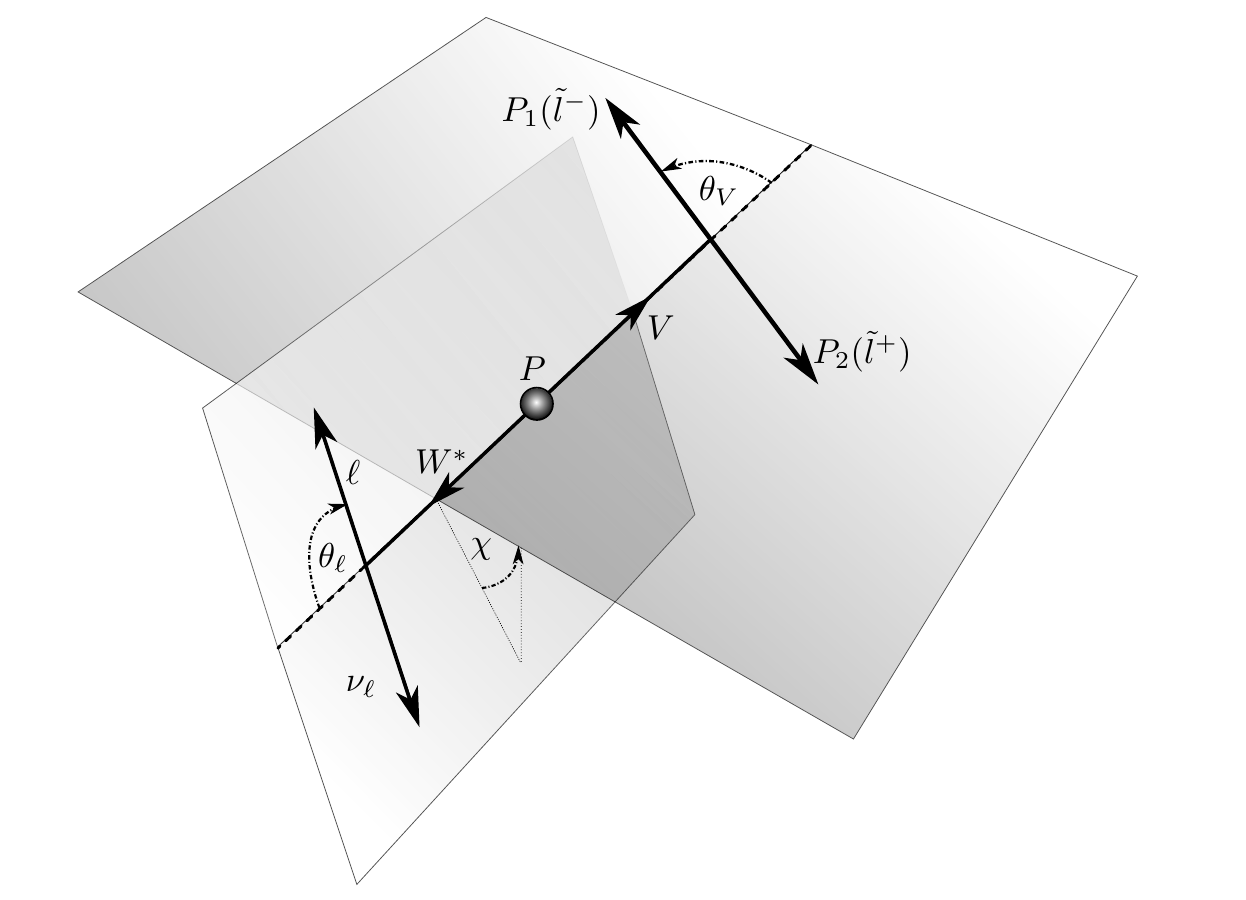}
\caption{Angle conventions for semileptonic decays of the form $P \!
\to \! V \ell \nu_\ell$, $V \! \to \! P_1 P_2$, where $P$ is a
pseudoscalar meson, $V$ is a vector meson, and $P_1$, $P_2$ ($\tilde
\ell^-$, $\tilde \ell^+$) are decay products of $V$.  In the first
relevant case described in the text, the decay chain is $B \! \to \!
D^* \ell \nu_\ell$, $D^* \! \to \! D \pi$.  In the case $B \! \to \!
J/\psi \, \ell \nu_\ell$, the labels $V \! \to \! \tilde
\ell^- \tilde \ell^+$ represent $J/\psi \! \to \! \mu^- \mu^+$.
\label{fig1}}\end{center}
\end{figure}

Consider first the semileptonic decay process $P \! \to \! V \ell
\nu_\ell$, where $P$ is a pseudoscalar meson decaying to a vector
meson $V$, which subsequently decays into a pseudoscalar meson pair
$P_1 P_2$ ({\it e.g.}, $B\rightarrow D^*\ell\nu_\ell$, $D^*\rightarrow
D\pi$).  The differential rate for such decays depends upon the
momentum transfer $q^2$ to the $\ell \nu_\ell$ pair and three angles:
$\theta_{V}$, the polar angle characterizing the direction of $P_1$
(measured in the $V$ rest frame) with respect to the direction of $V$
(measured in the $P$ rest frame); $\theta_\ell$, the polar angle
characterizing the direction of the lepton $\ell$ (measured in the
$W^*$ [virtual $W$] rest frame) with respect to the direction of $W^*$
(measured in the $P$ rest frame); and $\chi$, the azimuthal angle
between the $V P_1 P_2$ plane and the $W^* \ell \nu$ plane.  The
angles are shown in Fig.~\ref{fig1}, and agree with those defined in
Ref.~\cite{Richman:1995wm}.  A detailed description of how these
angles compare with other conventions in the literature appears in the
following section.

One defines the full 4-fold differential decay rate for this process:
${\frac{d\Gamma (P\to V \! \ell \nu_\ell, \, V \! \to P_1 P_2)}
{dq^2\, d \! \cos \theta_{\! V} d \! \cos \theta_\ell \, d\chi}}$.  We
frequently integrate over the three distinct angles, and therefore
introduce the collective symbol
\begin{equation} 
X^V_\ell \equiv \left\{ \cos \theta_\ell, \, \cos \theta_V, \, \chi
\right\} ,
\end{equation}
and define the integral measure over $X^V_\ell$ and the full
derivative with respect to $X^V_\ell$ as
\begin{equation}
\begin{split}
\int d X^V_\ell & \equiv \int_{-1}^{+1} d \! \cos \theta_\ell
\int_{-1}^{+1} d \! \cos \theta_{V} \! \!
\int_0^{2\pi} d \chi  \, , \\
\frac{d}{d X^V_\ell} & \equiv   \frac{d}
{d \! \cos \theta_{V} \, d \! \cos \theta_\ell \, d\chi} \, ,
\end{split}
\end{equation}
respectively.  Thus, the full differential cross section
${\frac{d\Gamma (P \to V \! \ell \nu_\ell, \, V \! \to P_1 P_2)} {dq^2
\, d \! \cos \theta_{\! V} \, d \! \cos \theta_\ell \, d\chi}}$ can be
denoted by $\frac{d\Gamma^V_{\ell}}{dq^2 \, d X^V_\ell }$, and the
total cross section by
\begin{equation} \label{eq:total}
\Gamma ^{V}_{\ell} = \int_{m_\ell^2}^{(M_P-M_V)^2} \!\! d q^2 \! \int
dX^V_\ell \, \frac{d\Gamma^V_\ell}{dq^2 \, d X^V_\ell } \, ,
\end{equation}
where $q^2$ is integrated over all kinematically allowed momentum
transfers, from the hadronic maximum recoil point $q^2 \! = \!
m_\ell^2$ (at which the $\ell$ is produced at rest in the $W^*$ rest
frame) to the hadronic zero-recoil point $q^2 \! = \! (M_P \! - \!
M_V)^2$ (at which the $V$ is produced at rest in the $P$ rest
frame).

Alternatively, consider a process in which the final-state hadron is a
weakly decaying pseudoscalar $P'$ ({\it e.g.}, $B\rightarrow D \ell
\nu_\ell$.  The kinematics is simpler because the $P'$ is a
(pseudo)scalar without strong decay modes.  The kinematical variables
are similar to those above (upon substituting $V \! \to \!  P'$), but
only the angle $\theta_\ell$ remains, and the full differential decay
rate is given by ${\frac{d\Gamma (P \to P' \ell \nu_\ell)} {dq^2 \, d
\! \cos \theta_\ell}}$.  For later compactness, let us define
$X^{P'}_\ell \! \equiv \! \left\{\cos\theta_\ell\right\}$.  The total
cross section is then
\begin{equation}\label{eq:totp}
\Gamma ^{P'}_\ell
= \int_{m_\ell^2}^{(M_P-M_{P'})^2 } \! \! d q^2 \! \int d
X^{P'}_\ell{\frac{d\Gamma^{P'}_\ell }
{dq^2 \,  d X^{P'}_\ell}} \, .
\end{equation}

One can trivially generalize Eqs.~(\ref{eq:total})--(\ref{eq:totp}) to
weighted cross sections $\Gamma^{H}_{\ell,i}$, where $H \! = \! V,P'$,
by integrating with a weight function $W_i(q^2,m_l^2,X^H_l)$:
\begin{equation}
\Gamma^{H}_{\ell,i} \equiv \int_{m_\tau^2}^{(M_P-M_H)^2 }
\! \! d q^2 \! \int \! d X^H_\ell\,
W_{i} \left({m_\ell^2},{q^2} \! , X^H_\ell \right)
\! \frac{d\Gamma^{H}_{\ell}}
{dq^2 \, d X^H_\ell } .
\end{equation}
Note that the $q^2$ bounds include only the allowable kinematic regime
for $\tau$ decays, independent of the lepton channel considered.  By
excluding the range $m_\mu^2 \! \le q^2 \! < \! m_\tau^2$, one ensures
that the same range of phase space is sampled in all channels.

With these definitions, one can construct ratios from
different combinations of $\ell$ and $W_i$.  The simplest of these,
$R^{H}_{i}$, are generalizations of the standard $R(H)$:
\begin{equation} \label{eq:test}
  R_{i}^{H} \equiv \frac{\Gamma^{H}_{\tau,i}}{\Gamma^{H}_{\mu,i}} .
\end{equation} 
Note that $q^2\geq m_\tau^2$ means the $R^{H}_{i}$ with $W_{i} \! =
\! 1$ are {\em not\/} the ratios $R(H)$ typically used in the
literature, which are instead defined as ratios of
the {\em full\/} decay widths to these lepton channels.

One has considerable freedom in choosing $W_{i}$, but not all choices
are useful.  For our purpose of removing form-factor and leptonic-mass
dependences, we initially restrict to forms in which $q^2$ and
$m_\ell^2$ only appear in the ratio
%
\begin{equation} \label{eq:epsdef}
\varepsilon \equiv \frac{m_\ell^2}{q^2} \, ,
\end{equation}
which always obeys $\varepsilon \leq 1$ in the allowed range for
$q^2$.  While $\varepsilon$ strictly depends upon $m_\ell$, we forgo
an index $\ell$ on $\varepsilon$ unless confusion would arise.

For decays $P \! \rightarrow \! P'$ ({\it e.g.}, $B\rightarrow
D\ell\nu_\ell$), one finds three $W_i(m_l^2,q^2,X^V_\ell)\equiv
W_i(\varepsilon,X^V_\ell)$ that remove the form-factor dependences
(their derivation appears below, in Sec.~\ref{sec:HA}, and they can be
recognized in Table~\ref{tab:Weight1b}):
\begin{equation} \label{eq:WDs}
\begin{split}
&W_a(\varepsilon,X^{P'}_\ell) = 
\frac{5  (-3 \cos^2 \! \theta_\ell + 1)}{2(1-\varepsilon)^3} \, ,  \\
&W_b(\varepsilon,X^{P'}_\ell) =
-\frac{\cos \theta_\ell }{\varepsilon (1-\varepsilon)^2} \,  ,\\
&W_c(\varepsilon,X^{P'}_\ell) = 
\frac{5 \cos^2 \! \theta_\ell - 1}
{\varepsilon (1-\varepsilon)^2} \, .
\end{split}
\end{equation}
Similarly, for decays to $V$ ({\it e.g.}, $B\rightarrow
D^*\ell\nu_\ell$, $D^*\rightarrow D\pi$), we construct form-factor
independent SM tests by choosing $W_{i}\left(m_\ell^2,q^2,X^V_\ell
\right) = W_{i} \left(\varepsilon, X^V_\ell \right)$ to be any of the
eight forms ({\it cf.} Table~\ref{tab:Weight1}):
\begin{equation} \label{eq:WD*s}
\begin{split}
& W_1(\varepsilon, X^V_\ell )  = \frac{(5 \cos^2 \! \theta_\ell - 1) 
(-5 \cos^2 \! \theta_{V} + 3) }{2(1-\varepsilon)^2} ,   \\
&W_2(\varepsilon, X^V_\ell ) =  \frac{5(-3 \cos^2 \! \theta_\ell + 1) 
(5 \cos^2 \! \theta_{V} - 1) }{4 (1-\varepsilon)^3} , \\
&W_3(\varepsilon, X^V_{\ell} )  =  \frac{\cos \theta_\ell 
(-5 \cos^2 \!\theta_{V}  + 3)}{(1-\varepsilon)^2} , \\
&W_4 (\varepsilon, X^V_\ell) =
\frac{25(\sin 2\theta_\ell \sin 2\theta_{V} \! \cos \chi)}
{4(1-\varepsilon)^3} , \\
&W_5(\varepsilon, X^V_\ell) =  -\frac{2  \cos 2\chi  }
{(1-\varepsilon)^3} , \\ 
&W_6(\varepsilon, X^V_\ell) =  -\frac{\cos \theta_\ell
(5 \cos^2 \!\theta_{V} - 1 )}{2\varepsilon (1-\varepsilon)^2} , \\
&W_7(\varepsilon, X^V_\ell)= \frac{(5 \cos^2 \! \theta_\ell - 1)
(5 \cos^2 \! \theta_{V} - 1)}{2 \varepsilon (1-\varepsilon)^2} ,  \\
& W_8(\varepsilon, X^V_\ell)  =\frac{ (-5 \! \cos^2 \! \theta_\ell + 2)
(-5 \cos^2 \! \theta_{V}+ 3) }{{ \varepsilon (1-\varepsilon)^2} }
\; .
\end{split}
\end{equation}

With these choices of $W_{i}$, by construction the SM predicts that
the ratios defined in Eq.~(\ref{eq:test}) satisfy
\begin{equation}
R_{i}^{H}=1 + {\cal O} (\alpha) \, ,
\label{eq:predict}
\end{equation}
where ${\cal O}(\alpha)$ indicates leading-order electroweak
corrections not included in our analysis, the same level currently
neglected in $R(H)$ calculations.  The prediction of
Eq.~(\ref{eq:predict}) for each $i$ can be viewed as a test of lepton
universality: Universality violations imply $R(H)$ generically differs
from unity.

At this stage, the angular and $\varepsilon$ factors appearing in
Eqs.~(\ref{eq:WDs})--(\ref{eq:WD*s}) seem quite arbitrary, and it may
seem unclear how they remove the form-factor dependence or should
yields $R_{i}^{h}=1$ in the SM\@.  In fact, the reason for both is
quite simple.  In Sec.~\ref{sec:HA}, the differential cross sections
are written in terms of helicity amplitudes (which are linear
combinations of the transition form factors).  It is shown below that,
when any $W_{i}$ given above is integrated over the differential cross
sections, one obtains a particular quadratic form of the helicity
amplitudes, for example:
\begin{equation}\label{eq:ex}
\begin{split}
&\frac{1}{G_0}\int d X^V_\ell \, W_{1}\left(\varepsilon, X^V_\ell
\right) \frac{d\Gamma^V_\ell }
{dq^2 \, d X^V_\ell }\\& = |H_+ (q^2)|^2 +  |H_- (q^2)|^2 \, ,
\end{split}
\end{equation}
where $H_+$ and $H_-$ are two helicity amplitudes defined in
Sec.~\ref{sec:HA}, and $G_0$ is a combination of overall fundamental
constants and known functions of $q^2$ (but not
$m_\ell^2$).\footnote{To be precise, $G_0$ is the coefficient
$\frac{d\Gamma_0}{dq^2}$ of Eq.~(\ref{eq:overall}) below, with the
factor $(1 \! - \! \varepsilon)^2$ removed.}  Furthermore, the $W_{i}$
are designed to remove the kinematic dependences on $\varepsilon$ such
that, for fixed $q^2$, the weighted differential cross section after
angular integration depends upon a fixed combination of helicity
amplitudes, independent of lepton flavor.  Therefore,
$\Gamma^V_{\ell,i}$ are integrals {\em only\/} of these special
combinations, so that, {\it e.g.}, Eq.~(\ref{eq:ex}) yields
\begin{equation}
R^V_1=\frac{\Gamma^V_{\tau,1}}{\Gamma^V_{\mu,1}}=
\frac{\int^{(M_P-M_V)^2}_{m_\tau^2}dq^2 \big( |H_+ (q^2)|^2 +  
|H_- (q^2)|^2 \big)}{\int^{(M_P-M_V)^2}_{m_\tau^2}dq^2
\big(|H_+ (q^2)|^2 +  |H_- (q^2)|^2 \big)}\,,
\end{equation}
which is manifestly unity in the SM, regardless of whether one can
determine the helicity amplitudes.  While one could compare different
lepton channels at the weighted differential cross-section level, such
analysis may be difficult because the data are sparse in some bins, or
the experimental analysis may not be straightforward for extracting
them.  Instead, by integrating in $q^2$, one can perform these
calculations on any data set that can produce $R(H)$, with improved
sampling statistics and reduced background for realistic experimental
situations.

One should note that while $W_a$, $W_c$, $W_1$, $W_2$, $W_7$, and
$W_8$ depend upon helicity-amplitude combinations appearing in the
total decay rates [see Eqs.~(\ref{eq:IntWidth}) and
(\ref{eq:IntWidth2})], $W_b$ and $W_{3-6}$ do not.  Therefore, to
explain the existing $R(H)$ tensions with BSM physics, these weights
are particularly important for the immediate analysis.  But tests
based upon $W_b$ and $W_{3-6}$ are interesting in their own right, as
they probe other aspects of possible SM violations.  These tests can
also be applied to $\overline{B}$ ($b \! \to \! c \, \ell^- \bar
\nu_\ell$) decays, using precisely the same $W_{i}$ except for an
overall sign change in $W_3$; but this sign is innocuous in $R^H_i$.

One is not restricted just to the weight functions
$W_{a,b,c}$ and $W_{1-8}$ discussed above.  Clearly,
any (possibly $q^2$-dependent) linear combinations of $W_{a,b,c}$ or
$W_{1-8}$ also yield valid weight functions $W$ for which the SM
predictions of Eq.~(\ref{eq:predict}) hold:
\begin{equation}
W \! \left(m_\ell^2,q^2 \! ,X^H_\ell \right) \equiv \sum_{j}
f_j(q^2) W_j \! \left(\varepsilon, X^H_\ell \right) \, , \\
\label{eq:Wls}
\end{equation}
where $j$ is the set of allowed weight functions for the $H$ decay
channel, either $a,b,c$ for $H \! = \! P'$, or $1-8$ for $H \! = \!
V$, and $f_j(q^2)$ are functions of $q^2$ that are independent of
lepton flavor.  One would be mistaken to presume these linear
combinations provide no new information.  First, the functions $f$ can
be chosen to emphasize different $q^2$ regions, as opposed to using an
unweighted $q^2$ integral.  When using experimental results, it may be
advantageous to choose $f$ to reduce the experimental uncertainties in
the ratios by choosing linear combinations of weight functions or
their coefficients in Eq.~(\ref{eq:Wls}) that minimize the
contribution from kinematical regions with larger uncertainties, {\it
e.g.}, close to the $q^2$ minimum value of $m_\tau^2$.  Second, even
for $f$ constant, the ratio of averages using Eq.~(\ref{eq:Wls})
would include terms containing ratios of the form $W_j/W_k$ where $j
\! \ne \! k$, which are absent from ratios containing a single
weight function.  In short, the ratio of sums differs from the sum of
ratios.

It is straightforward to test these relations experimentally.
Consider an idealized experimental situation: One has an arbitrarily
large amount of data in a complete set of $N_{H}$ decay events, of
which $N_{H\ell}$ are semileptonic decay events in the $\ell=\mu,\tau$
channels; the momentum transfer and the angles are measured to
arbitrary accuracy; and for each such event $j$ with precisely
determined kinematics, one can determine two probabilities to
arbitrary accuracy: the probability $P^{\rm b}_j$ that an event with
kinematics $j$, which has been identified as a possible $P\rightarrow
H$ decay, is actually a background event (rather than being a true
decay, which has probability $\bar P^{\rm b}_j \!  = \! 1 \! - \!
P^{\rm b}_j$), and the probability $P^{\rm d}_j$ is measured and
correctly identified ({\it i.e.}, the total efficiency for detection
and identification is known).

In such a case, the statistical average of ratios $R_{i}^{H}$ can be
determined experimentally by
%
\begin{equation} \label{eq:RWexpt} \begin{split}
&\langle R_{i}^H \rangle  = \frac{
\sum_{j=1}^{N_{\! H \tau}} \frac{\bar P_j^{\rm b}}
{P_j^{\rm d}} W_{i}(m^2_\tau, q_j^2, X^H_{\tau,j})}{ 
\sum_{j'=1}^{N_{H\mu}} \frac{\bar P_{j'}^{\rm b}}
{P_{j'}^{\rm d}}
W_{\! i}(m^2_\mu, q_{j'}^2, X^H_{\mu,j'})
\Theta (q_{j'}^2 -m_\tau^2)
} ,
\end{split}\end{equation}
%
where the brackets indicate a statistical average for the quantity,
and the index $j$ ($j'$) indicates a particular decay event in the
$\ell \! = \! \tau$ ($\ell \! = \! \mu$) channel.  $\Theta$ denotes a
Heaviside step function that ensures the sums cover the same kinematic
region in $q^2$.  Equation~(\ref{eq:RWexpt}) represents a pure
counting experiment: Since the events in both the numerator and
denominator are sampled probabilistically, they effectively map out
the $\tau$ and $\mu$ differential decay-width distributions; by
weighting each event with the appropriate function $W_{i}$, one
develops an approximation to the relations of Eq.~(\ref{eq:predict}).

A few comments about the experimental implementation of
Eq.~(\ref{eq:RWexpt}) is in order.  First, one can in principle obtain
reliable estimates of $R_{i}^{H}$ (for at least some choices of the
weight functions $W_{i}$) with far less data than is needed to extract
the form factors.  In particular, one does not need the full angular
dependence of the data at identical values of $q^2$ to obtain
well-converged sums in Eq.~(\ref{eq:RWexpt}).  In this sense, the
situation is similar to the extraction of $R(H)$ in
Refs.~\cite{Lees:2012xj,Lees:2013uzd,Huschle:2015rga,Sato:2016svk,
Aaij:2015yra,Hirose:2016wfn,Wormser:2017hsx,Aaij:2017tyk,
Hirose:2017dxl}.

Second, while theoretically $R_{i}^{H}$ do not depend upon knowledge
of the form factors, the experimental extractions of the ratios {\em
can\/} depend upon the form factors, to the extent that they are used
in the determination of $P_{j,j'}^{\rm b}$ and $P_{j,j'}^{\rm d}$
(which is a potential major concern, as the experimental uncertainty
on $R(J/\Psi)$ is dominated by form-factor uncertainties used to
discriminate backgrounds~\cite{Aaij:2017tyk}).

Third, throughout our analysis we assume that the $\tau$ can be fully
reconstructed.  In practice, such detailed information might not be
accessible, in which case one could either generalize the technique
presented here by including the angular dependences from the $\tau$
decay products, or restrict to a set of $W_i$ that can be reliably
extracted.  The latter approach is considered in
Ref.~\cite{Alonso:2017ktd}, where the authors study the restricted set
of useful observables when only limited information can be extracted
from the final states of $\tau$ decays.

Fourth, in principle an infinite number of $R_i^H$ exist, due to the
arbitrary linear combinations and coefficient $q^2$ dependences
allowed by Eq.~(\ref{eq:Wls}).  One thus obtains an infinite number of
tests of the SM\@.  One can exploit this freedom in two complementary
ways.  First, if one believes that the discrepancies are hints of a
particular BSM model, one can choose $W_i$ to maximize sensitivity to
those particular violations.  Alternately, one may exploit the freedom
in choosing $W_i$ to reduce the experimental uncertainties by choosing
linear combinations in Eq.~(\ref{eq:Wls}) that minimize the
contribution from kinematical regions with larger uncertainties, {\it
e.g.}, the limit $\varepsilon \! \to \! 1$ ($q^2 \! \to \! m_\ell^2$)
where fewer events should occur, and therefore which are very
sensitive to statistical fluctuations.
 
In this context, it is worth noting that all $W_i$ have a coefficient
as $\varepsilon \! \to \! 1$ ($q^2 \! \to \! m_\ell^2$) at least as
singular as $(1-\varepsilon)^{-2}$, which compensates for a factor of
$(1-\varepsilon)^{2}$ in the total cross section arising from phase
space and helicity suppression constraints.  In Eq.~(\ref{eq:test}),
these factors cancel and yield finite results.  However, in
an experimental situation, the data in this region can become
particularly sensitive to statistical fluctuations since there should
be fewer events in the $\tau$ channel.\footnote{Owing to the cutoff
$q^2 \! \ge \! m_\tau^2$ in Eq.~(\ref{eq:test}), the factor $(1 \! -
\! \varepsilon)^{-2}$ in the $\mu$ channel is always within 1\% of
unity.}  To remove this sensitivity, one may exploit the freedom in
choosing the functions $f$ in Eq.~(\ref{eq:Wls}) to ensure that they
go to zero as $q^2 \! \rightarrow \!  m_\tau^2$, and thereby suppress
large fluctuations.  This freedom is particularly important for $W_a$,
$W_2$, $W_4$, and $W_5$, which scale as $(1 \! - \!
\varepsilon)^{-3}$.

Similarly, $W_b$, $W_c$, $W_6$, $W_7$, and $W_8$ contain overall
factors of $1/\varepsilon$.  For the $\mu$ channel, this factor is
always quite large---at least 280.  These factors arise in
helicity-suppressed helicity amplitudes in the differential cross
section.  It will therefore likely be difficult to extract these
amplitudes accurately, since statistical or systematic errors can
swamp the data.  Thus, the most robust tests of the SM avoid reliance
on these $W_i$.  However, BSM models could enhance these amplitudes
such that deviations from the SM predictions might be large enough to
tease out using linear combinations containing these weight functions.

We identify another class of SM tests for $P\rightarrow V$ that is not
sensitive to violations of lepton universality, but rather probes
other aspects of the SM while remaining independent of the form
factors.  This class of test also depends upon ratios of two weight
functions, but only a single lepton flavor.  These tests reflect the
nature of the weight functions $W_1$ and $W_8$, which have two
distinct angular dependences, and yet yield the same the helicity
amplitude combinations as in Eq.~(\ref{eq:ex}):
\begin{equation}
\begin{split}
\label{eq:rvel}
&R^V_{\ell,nd} \equiv  \frac{\int_{m_\ell^2}^{(M_P-M_V)^2 }
d q^2 \! \int \! dX^V_\ell \, W_n(m_\ell, q^2 \! , X^V_\ell)
\frac{d\Gamma^V_\ell}
{dq^2 \, d X^V_\ell}}{\int_{m_\ell^2}^{(M_P-M_V)^2 } d q^2 \!
\int \! dX^V_\ell \, W_d(m_\ell, q^2 \! ,X^V_\ell)
\frac{d\Gamma^V_\ell}
{dq^2 \, d X^V_\ell}} \, ,
\end{split}
\end{equation}
with the weight functions $W_{n,d}$ defined by
\begin{eqnarray}
\lefteqn{W_{i} (m^2_\ell, q^2 \! , X^V_\ell)=}& & \nonumber \\
&\phantom{xx} & h(q^2) \left[ \cos^2 \! \phi_i(q^2) \,
W_1(\varepsilon\! ,X_\ell) +\sin^2 \! \phi_i(q^2) \,
W_8(\varepsilon \! ,X_\ell) \right] \, , \nonumber \\
\label{eq:Weight18}
\end{eqnarray}
where $h(q^2)$ and $\phi_i(q^2)$ are specified functions of $q^2$.
The SM prediction is again $R^V_{\ell,nd} =1 + {\cal O} (\alpha)$ for
both $\ell=\mu,\tau$, and any choice of $h(q^2)$, $\phi_n(q^2)$, and
$\phi_d(q^2)$.

Since this test depends upon $W_8$, which has a coefficient
$1/\varepsilon$ that is large over much of the kinematic region, a
useful test will likely select functions $\phi(q^2)$ that deemphasize
the region where $\varepsilon$ is especially small.  Note that, since
the ratios $R^V_{\ell,nd}$ refer to a single species of lepton $\ell$,
the integrations in both the numerator and denominator extend to
$\varepsilon \! = \! 1$, unlike $R^H_{i}$, which is restricted to
$q^2\geq m_\tau^2$.

Having shown how to construct tests of the SM from the weight
functions $W_i$, in the next section we demonstrate how these $W_i$
arise naturally in association with the helicity amplitudes appearing
in the decay rates.

\section{Helicity Amplitudes}
\label{sec:HA}

\subsection{The Decays $P \! \to \! V \ell \nu_\ell, \ V \! \to \! P_1
P_2$}

The form factors for the transition of a pseudoscalar meson $P$ (mass
$M$, momentum $p$) to a vector meson $V$ (mass $m$, momentum $p'$,
polarization vector $\epsilon$) are defined as~\cite{Boyd:1997kz}
\begin{eqnarray}
\left< V(p', \epsilon) \left| V^\mu \right| P(p) \right> & = & i g(q^2)
\epsilon^{\mu \alpha \beta \gamma} \epsilon^*_\alpha p'_\beta p_\gamma
\, , \nonumber \\
\left< V(p', \epsilon) \left| A^\mu \right| P(p) \right> & = & f(q^2)
\epsilon^{*\mu}  + ( \epsilon^* \cdot p )
\left[ a_+ (q^2) ( p + p' )^\mu \right. \nonumber \\ & &
\left. \hspace{2em} + a_- (q^2) ( p - p' )^\mu \right] ,
\end{eqnarray}
where the momentum transfer is given by $q^2 \! \equiv \! (p - p')^2$.
The first calculations of the complete differential decay rates of the
semileptonic process $P \! \to \! V \ell \nu$, $V \! \to \! P_1 P_2$
including finite charged-lepton mass effects appeared in
Refs.~\cite{Korner:1989qb,Gilman:1989uy}.  The helicity amplitudes
defined in the classic review Ref.~\cite{Richman:1995wm} and still
commonly used ({\it e.g.}, by the Belle
Collaboration~\cite{Abdesselam:2017kjf}) are given by
\begin{eqnarray}
H_\pm (q^2) & = & -H_\pm^{\rm KS} = -f \pm M p^{\vphantom\dagger}_V g
\, , \nonumber \\
H_0 (q^2) & = & -H_0^{\rm KS} = -\frac{1}{\sqrt{q^2}} {\cal F}_1
\nonumber \\ & = & -\frac{1}{2m\sqrt{q^2}}
\left[ (M^2 \! - m^2 \! - q^2) f + 4M^2 p^2_V \, a_+ \right] \, ,
\nonumber\\
H_t & = & -H_t^{\rm KS} = -\frac{M p^{\vphantom\dagger}_V}{\sqrt{q^2}}
{\cal F}_2 \nonumber \\ & = &
-\frac{M p^{\vphantom\dagger}_V}{m\sqrt{q^2}}
\left[ f + (M^2 \! - m^2) a_+ + q^2 a_- \right] \, .
\end{eqnarray}
Here, $p^{\vphantom\dagger}_V$ is the momentum magnitude of the $V$
(or virtual $W$) in the center-of-momentum (c.m.) frame of $P$:

\begin{equation} \label{eq:pV}
p^{\vphantom\dagger}_V \equiv \sqrt{ \frac{[q^2 - (M+m)^2] [q^2 -
    (M-m)^2]} {4M^2} } \, .
\end{equation}
The subscript on $H$ gives the $W^*$ helicity: $\pm 1$ and 0 for
$J_{W^*} \! = \! 1$, $t$ (timelike) for $J_{W^*} \! = \! 0$.  The
superscript KS indicates the notation of
Ref.~\cite{Korner:1989qb},\footnote{Although
Ref.~\cite{Richman:1995wm} does not define $H_t$, it is natural to
extrapolate from Ref.~\cite{Korner:1989qb}, using the same relative
sign as for $H_{\pm, 0}$.} and the combinations ${\cal F}_{1,2}$ are
those defined in Ref.~\cite{Boyd:1997kz}.  The precise number of
independent helicity amplitudes for semileptonic processes
is most easily computed by considering the crossed process with
all hadrons in the initial state and all leptons in the final state,
and then imposing assumed conservation laws ({\it e.g.}, CP
conservation) on the system~\cite{Ji:2000id,Lebed:2001ic}.

The full 4-fold differential cross section for the semileptonic decay
$P (Q\bar q) \! \to \! V (q'
\bar q) \, \ell \nu_\ell$, $V \! \to \! P_1 P_2$, reads
\begin{widetext}
\begin{align}
&\frac{d\Gamma (P \! \to V \! \ell \nu_\ell, \, V \! \to P_1
P_2)} {dq^2 \, d \! \cos \theta_V \, d \! \cos \theta_\ell \, d\chi}
=\frac{3}{8(4 \pi)^4} G_F^2 \left| V_{q' Q} \right|^2
\frac{p^{\vphantom\dagger}_V q^2 (1-\varepsilon)^2}{M^2} {\cal B}
(V \to P_1 P_2)\nonumber
\\&\phantom{xxx}\times\left\{ \;
\left[ ( 1 - \eta \cos \theta_\ell )^2  + \varepsilon \sin^2 \!
\theta_\ell \right] \sin^2 \! \theta_V \left| H_+ (q^2) \right|^2 +
\left[ ( 1 + \eta \cos \theta_\ell )^2 + \varepsilon \sin^2 \!
\theta_\ell \right]  \sin^2 \! \theta_V \left| H_- (q^2) \right|^2
\right. \nonumber\\ 
&\phantom{xxx}+ \, 4 \left( \sin^2 \! \theta_\ell
+ \varepsilon \cos^2 \! \theta_\ell \right) \cos^2 \! \theta_V
\left| H_0 (q^2) \right|^2 - 2 \eta \sin \theta_\ell \sin 2\theta_V \!
\cos \chi \left\{\left[ 1 - (1 - \varepsilon) \eta \cos \theta_\ell
\right] {\rm Re} \, H_+ H^*_0 (q^2)\right.\nonumber\\
&\phantom{xxx}-\left. \left[ 1 + (1 - \varepsilon) \eta \cos
\theta_\ell \right]
{\rm Re} \, H_- H^*_0 (q^2) \right\}
- 2 \sin^2 \! \theta_\ell \sin^2 \! \theta_V \cos 2\chi \,
( 1 - \varepsilon ) \, {\rm Re} \, H_+ H^*_- (q^2) \nonumber \\
& \phantom{xxx} + \, 4\varepsilon \! \,\left[ \cos^2 \! \theta_V
\left| H_t (q^2) \right|^2  
- \! 2 \cos \theta_\ell \cos^2 \! \theta_V \,
{\rm Re} \, H_0 H^*_t (q^2) \left.
+ \sin \theta_\ell \sin 2\theta_V \cos \chi \, \frac 1 2 {\rm Re}
\left( H_+ \! + H_- \right) \! H^*_t \! (q^2) \right] \right\}  ,
\label{eq:RBwidth}
\end{align}
\end{widetext}
where $q^2$ is the momentum transfer (or equivalently, the invariant
squared mass of the $W^*$), and $\eta = \pm 1$ corresponds to
processes with lepton pairs $\ell^- \bar \nu_\ell$ and $\ell^+
\nu_\ell$, respectively ({\it i.e.}, twice the neutrino helicity).

This expression is equivalent to Eq.~(22) in~\cite{Korner:1989qb} if
one replaces $\theta_{\rm KS} \! = \! \pi \! - \! \theta_\ell$.  In a
conventional calculation, the angular factors emerge from choosing a
helicity basis of polarization vectors $\epsilon$ for $V$ and
$\epsilon_W$ for $W^*$, and the lepton 4-momenta $p_\ell$ and $p_\nu$.
More generally, they are Wigner rotation matrices connecting various
helicity states; adapting from Ref.~\cite{Dey:2015rqa}, one may write
\begin{eqnarray}
\lefteqn{\frac{d\Gamma (P \! \to V \! \ell \nu_\ell, \, V \! \to P_1
P_2)} {dq^2 \, d \! \cos \theta_V \, d \! \cos \theta_\ell \, d\chi}}
& & \nonumber \\
& = & \frac{G_F^2 |V_{q' Q}|^2 q^2 (1 - \varepsilon)^2
{\cal B} (V \to P_1 P_2)} {M^2 (4\pi)^4} \nonumber \\ & \times &
\!\!\!\! \sum_{\kappa = \eta, 0} \left| \sum_{\substack{\lambda = 0,
\pm 1 \\ J = 0, 1}} \!\! \sqrt{2J \! + 1} \, (-1)^J
{\cal H}^J_{\lambda, \kappa} d^J_{\lambda, \kappa} (\theta_\ell)
d^1_{\lambda, 0} (\theta_V \! ) e^{i\lambda \chi} \right|^2 \! .
\nonumber \\ \label{eq:Wigner}
\end{eqnarray}
Unlike in Ref.~\cite{Dey:2015rqa}, the $V$ spin in this expression is
fixed to 1; and the $W^*$ spin $J$ is no longer limited just to 1, but
is also allowed to assume the ($J \! = \! 0$) timelike polarization
$\epsilon_W^\mu \! = \! q^\mu / \sqrt{q^2}$.  When $q^\mu
\! = \!  p_\ell^\mu + p_\nu^\mu$ is contracted with the lepton
bilinear, {\it e.g.}, $\bar u (p_\ell) \gamma^\mu v_L (p_\nu)$ or
$\bar v_R (p_\nu) \gamma^\mu u (p_\ell)$ in the case $\eta \! = \!
+1$, use of the Dirac equation produces an overall coefficient of
$m_\ell / \sqrt{q^2}$ in the amplitude.  The total lepton helicity
$\kappa$ in the $W^*$ rest frame is given by $\kappa
\! = \! \lambda_\ell + \eta/2$ and equals $\eta$ for the spin
non-flip transition (right-handed $\bar \nu$ and left-handed $\ell^-$
for $\eta \! = \!  +1$, left-handed $\nu$ and right-handed $\ell^+$
for $\eta \!  = \! -1$) and 0 for the spin-flip transition (opposite
helicities for $\ell$).  The spin non-flip transition gives the
leading-order amplitude in the $V \! - \! A$ theory, which in the
$W^*$ rest frame gives a contribution to the rate proportional to
$2p_\ell (E_\ell \! + \!  p_\ell) = \!  q^2 \! - m_\ell^2$, while the
spin-flip contribution is proportional to $2p_\ell (E_\ell \! - \!
p_\ell) = \! (q^2 \! - m_\ell^2) (m_\ell^2/q^2)$.  The lepton mass
parameter $\varepsilon$ thus appears in four places in the
differential rate: (i) in the quasi-two-body phase space factor
$p_\ell \propto q^2 \! - m_\ell^2$ in $W^* \to \ell \nu$; (ii) in the
factor $p_\ell$ common to both spin non-flip and spin-flip transitions
in $V \! - \! A$ theory; (iii) in the additional suppression of
spin-flip transitions in the $V \! - \!  A$ theory; and (iv) in the
coupling of a timelike $W^*$ in any vectorlike theory.  A pedagogical
review of these points appears in Ref.~\cite{Korner:2014bca}.

The amplitudes ${\cal H}^J_{\lambda, \kappa}$ in Eq.~(\ref{eq:Wigner})
incorporate the nonperturbative physics in terms of helicity
amplitudes (and ultimately, form factors), while the Wigner rotation
matrices $D^J_{m' \! ,m} (\alpha, \beta, \gamma) \! = \! e^{-im' \!
\alpha} d^J_{m' \! ,m} (\beta) e^{-im\gamma}$ encapsulate all the
nontrivial angular correlations.  Only one azimuthal angle $\chi$ is
required to describe the decay, which is that of the $D^* \! \to \!
D\pi$ decay plane with respect to the $W^* \! \to \ell \nu$ decay
plane (Fig.~\ref{fig1}).  The factor $(-1)^J$ represents the sign
difference in the norm between timelike and spacelike $W^*$
polarizations.  The sums are further restricted by the factor
$d^J_{\lambda, \kappa}$ when $J \! = \! 0$ to have $\lambda \!  = \!
\kappa \! = \! 0$.  Lastly, note the great simplification due to the
decay of the spin-1 $V$ to spinless particles $P_{1,2}$: Only the
matrices $d^1_{\lambda, 0}$ are needed to describe the angular
dependence for that subprocess.

The precise definitions of the angles are depicted in Fig.~\ref{fig1}
and agree with those in Ref.~\cite{Richman:1995wm}: Starting with the
rest frame of the spinless $P$, the $V$-$W^*$ decay axis is identified
with the $z$-axis, {\it i.e.}, ${\bf p}^{\vphantom\dagger}_V \! = \!
+\hat{\bf z}$.  Then the helicity $\lambda \equiv
\lambda^{\vphantom\dagger}_V \! = \! \lambda_{W^*}$.  Boosting into
the $W^*$ rest frame, one finds the $\ell$ and $\nu$ back-to-back, and
defines $\theta_\ell$ as the polar angle of $\ell$ with respect to the
$W^*$ direction as measured in the $P$ rest frame.  Similarly,
boosting into the $V$ rest frame, one finds $P_1$ and $P_2$
back-to-back, and defines $\theta^{\vphantom\dagger}_V$ as the polar
angle of $P_1$ (which we take as the heavier of $P_{1,2}$, such as $D$
in $D^* \! \to \! D\pi$) with respect to the $V$ direction as measured
in the $P$ rest frame.  Finally, we take $\chi$ as the azimuthal angle
of the $V P_1 P_2$ plane with respect to the $W^* \ell \nu$ plane; to
be precise, Refs.~\cite{Richman:1995wm,Abdesselam:2017kjf} actually
exhibit $\chi$ as the {\em clockwise\/} rotation of the $V P_1 P_2$
plane with respect to the $W^* \ell \nu$ plane, as viewed with respect
to the axis ${\bf p}^{\vphantom\dagger}_V \! = \! +\hat{\bf z}$, which
explains the relative sign of the phase in Eq.~(\ref{eq:Wigner})
compared to that in the conventional notation given
above.\footnote{Strictly speaking, this $\chi$ differs from the one
($\chi^{\rm KS}$) used in Ref.~\cite{Korner:1989qb} by $\chi \! = \!
- \! \chi^{\rm KS}$.  Furthermore, a reanalysis of $\chi^{\rm Dey}$
used in Ref.~\cite{Dey:2015rqa} shows that $\chi \! = \! \pi \! + \!
\chi^{\rm Dey}$ : To obtain Eq.~(\ref{eq:RBwidth}), the factor $e^{i
\lambda \chi}$ in Eq.~(\ref{eq:Wigner}) must be replaced with
$e^{i \lambda (\pi + \chi)}$.}

Once the amplitudes ${\cal H}^1_{\lambda, |\kappa|=1} = H_\lambda$,
${\cal H}^1_{\lambda, 0} = \sqrt{\varepsilon/2} H_\lambda$, and ${\cal
H}^0_{0,0} = \sqrt{3\varepsilon/2} H_t$ are inserted and all
CP-violating terms (those proportional to the imaginary parts of
interference terms, ${\rm Im} \, H_i H^*_j$, and hence proportional to
$\sin \chi$) are neglected, one obtains Eq.~(\ref{eq:RBwidth}).
Retaining CP violation modifies Eq.~(\ref{eq:RBwidth}) in such a way
that, for each term of the form $\cos (n\chi) \, {\rm Re} \, H_i
H^*_j$, where $n \! = \! 1$ or 2 and $i \! \ne \! j$, one introduces
an additional term of the form $\pm \sin (n\chi) \, {\rm Re} \, H_i
H^*_j$, in which the sign depends upon the particular amplitudes
$H_{i,j}$. Such effects appear in the analysis of
Ref.~\cite{Dey:2015rqa} and are relevant to studies such as in
Ref.~\cite{Aloni:2018ipm}.

The question now becomes whether one can extract independently the
helicity amplitude combination ${\rm Re} \, H_i H_j^*$ from each term
in Eq.~(\ref{eq:RBwidth}), and indeed, since most of the
$\varepsilon$-suppressed terms also carry distinct angular dependence,
the combinations $\varepsilon {\rm Re} \, H_i H_j^*$ as well.  Of the
15 such terms in Eq.~(\ref{eq:RBwidth}), some are clearly linearly
dependent: For example, there is no way to extract the difference
between $\varepsilon |H_+|^2$ and $\varepsilon |H_-|^2$, nor ${\rm Re}
\, H_+ H^*_-$ independently of $\varepsilon {\rm Re} \, H_+ H^*_-$.
This linear dependence arises partly through the restrictive form of
the $V \! - \! A$ interaction and partly through the simplicity of the
helicity structures appearing in $V \! \to \! P_1 P_2$.  As for the
remaining terms, one might think to use the orthonormality of $D$
matrices, first reducing pairs of the matrices via the Clebsch-Gordan
series
\begin{eqnarray}
\lefteqn{D^j_{mk} (\alpha, \beta, \gamma) D^{j'}_{m'k'} (\alpha,
\beta, \gamma)} & & \nonumber \\
& = & \sum_{J=|j-j'|}^{j+j'} \left< j m \, j' \! m' \left| \right. \!
J (m+m') \right> \left< j k \, j' \! k' \left| \right. \! J (k+k')
\right> \nonumber \\
& & \times D^J_{(m+m')(k+k')} (\alpha, \beta, \gamma) \, .
\end{eqnarray}
While this method identifies the linearly dependent terms, a much
simpler approach is available for Eq.~(\ref{eq:RBwidth}): By
inspection, one first separates terms with $\chi$ dependence into the
sets 1, $\cos \chi$, and $\cos 2\chi$, which are clearly independent
by Fourier analysis.  Of these, the $\cos 2\chi$ term in
Eq.~(\ref{eq:RBwidth}) is unique, while the only independent
structures multiplying $\cos \chi$ are clearly $\sin
\theta_\ell \sin 2\theta_V$ and $\sin 2\theta_\ell \sin 2\theta_V$.
Of the $\chi$-independent terms, the independent $\theta_\ell$
structures are $\cos \theta_\ell$, $\cos^2 \theta_\ell$, and $\sin^2
\theta_\ell$.  The corresponding independent $\theta_V$ structures can
always be reduced to the set $\cos^2 \theta_V$ and $\sin^2 \theta_V$,
so that Eq.~(\ref{eq:RBwidth}) contains 6 linearly independent
$\chi$-independent terms.  In total, exactly 9 structures in
Eq.~(\ref{eq:RBwidth}) are independent.

One can further extract the coefficient of each angular structure
using orthogonality almost by inspection: For example, a term
proportional to $\sin \theta_\ell \sin 2\theta_V \cos \chi$ is most
easily separated from all other structures present simply by
integrating with the weight function
\begin{equation}
\int_{-1}^{+1} d \! \cos \theta_\ell \, \sin \theta_\ell
\int_{-1}^{+1} d \! \cos \theta_V \sin 2\theta_V \! \int_0^{2\pi}
d \chi \cos \chi \, .
\end{equation}
Defining an overall differential width coefficient,
\begin{equation}
\frac{d\Gamma_0}{dq^2} \equiv \frac{ G_F^2 \left| V_{q' Q}
\right|^2}{96\pi^3} \frac{p^{\vphantom\dagger}_V q^2
(1-\varepsilon)^2}{M^2} {\cal B} (V \to P_1 P_2) \, ,
\label{eq:overall}
\end{equation}
which is $64\pi/9$ times the coefficient in the first line of
Eq.~(\ref{eq:RBwidth}), one extracts helicity amplitude combinations
by performing the integrals
\begin{eqnarray} \label{eq:Weight}
& & \left( \frac{d\Gamma_0}{dq^2} \right)^{\! -1} \! \! \int_{-1}^{+1}
d \! \cos \theta_\ell \int_{-1}^{+1} d \! \cos \theta_V \! \!
\int_0^{2\pi} \! d \chi \, w_0(\theta_\ell, \theta_V \! , \chi)
\nonumber \\ & & \times
\frac{d\Gamma}{dq^2 \, d \! \cos \theta_V \, d \! \cos \theta_\ell \,
d\chi} \, ;
\end{eqnarray}
the required weight functions $w_0(\theta_\ell, \theta_V \! , \chi)$
and the 9 independent simple combinations of helicity amplitudes that
can be extracted are listed in Table~\ref{tab:Weight1}.  The full
differential width $d\Gamma/dq^2$ is of course obtained simply by
setting $w_0 \! = \! 1$, and reads
\begin{eqnarray}
\frac{d\Gamma}{dq^2} & = & \frac{d\Gamma_0}{dq^2} \left\{
\left( 1 + \frac{\varepsilon}{2} \right)
\left( |H_+|^2 + |H_-|^2 + |H_0|^2 \right) \right. \nonumber \\
& & \hspace{2em} + \left. \frac 3 2 \varepsilon |H_t|^2 \right\} \, .
\label{eq:IntWidth}
\end{eqnarray}

The results of this analysis identify several interesting features:
First, the squared amplitudes $|H_\pm|^2$ are the only ones that can
be extracted independently of the lepton mass correction
$\varepsilon$; indeed, $H_t$ is always accompanied by a factor
$\varepsilon$, and its mixing with $H_0$ prevents an
$\varepsilon$-independent determination of $|H_0|^2$.  Perhaps most
interesting from the point of view of lepton universality studies is
that the ratio of the eighth line of Table~\ref{tab:Weight1} to the
first, whose integrals differ only in the $\theta_\ell$ weighting,
gives a unique determination of the lepton mass parameter
$\varepsilon$.  To be explicit, first integrate to obtain
\begin{eqnarray}
\frac{d\Gamma_1}{dq^2 \, d \! \cos \theta_\ell} & \equiv &
\int_{-1}^{+1} d \! \cos \theta_V ( -5\cos^2 \! \theta_V + 3 )
\nonumber \\ & \times & \int_0^{2\pi} \! d \chi \, \frac{d\Gamma}
{dq^2 \, d \! \cos \theta_V \, d \! \cos \theta_\ell \, d\chi} \, ,
\label{eq:extracte1}
\end{eqnarray}
which is {\em not\/} the same as $d\Gamma/dq^2 d\cos \theta_\ell$, due
to the presence of the extra $\theta_V$-dependent term.  Then one
finds
\begin{equation}
\varepsilon = \frac{\int_{-1}^{+1} d \! \cos\theta_\ell \, \frac 1 2
(5 \! \cos^2 \! \theta_\ell - 1) \, \frac{d\Gamma_1}{dq^2 d \!
\cos \theta_\ell}} {\int_{-1}^{+1} d \! \cos\theta_\ell \, (-5 \!
\cos^2 \! \theta_\ell + 2) \, \frac{d\Gamma_1}{dq^2 d \! \cos
\theta_\ell}}
\, . \label{eq:extracte2}
\end{equation}
The same relations have been used to a rather different effect in
Eqs.~(\ref{eq:rvel})--(\ref{eq:Weight18}).

\begin{table*}[ht]
\setlength\extrarowheight{2.5pt}
\caption{Weight functions $w_0(\theta_\ell, \theta_V \! , \chi)$
integrated against the full 4-fold differential width
Eq.~(\ref{eq:RBwidth}) for processes $P \! \to V \! \ell \nu_\ell, \,
V \! \to P_1 P_2$ in the manner described in Eq.~(\ref{eq:Weight}).
They apply to cases where $V$ decays to a state of total
spin-projection zero along the decay axis.}
\label{tab:Weight1}
\begin{tabular}{c c}

$w_0(\theta_\ell, \theta_V \! , \chi)$ & {\rm Extracted
helicity amplitude} \\
\hline
$\frac 1 2 (5 \cos^2 \! \theta_\ell - 1) (-5 \cos^2 \! \theta_V + 3)$
& $|H_+|^2 \! + |H_-|^2$ \\
$\frac 5 4 (-3 \cos^2 \! \theta_\ell + 1) (5 \cos^2 \! \theta_V - 1)$
& $|H_0|^2 ( 1 - \varepsilon )$ \\
$-\eta \cos \theta_\ell (-5 \cos^2 \! \theta_V + 3)$ &
$|H_+|^2 \! - |H_-|^2$ \\
$\frac{25}{4} \sin 2\theta_\ell \sin 2\theta_V \cos \chi$ &
$(1 - \varepsilon) {\rm Re} \, (H_+ \! + H_-) H^*_0$ \\
$-2 \cos 2\chi$ & $(1 - \varepsilon) {\rm Re} \, H_+ H^*_-$ \\
$-\frac 1 2 \cos \theta_\ell (5 \cos^2 \! \theta_V - 1)$ &
$\varepsilon \, {\rm Re} \, H_0 H^*_t$ \\
$\frac 1 2 (5 \cos^2 \! \theta_\ell - 1) (5 \cos^2 \! \theta_V - 1)$ &
$\varepsilon \left( |H_0|^2 + |H_t|^2 \right)$ \\
$(-5 \! \cos^2 \! \theta_\ell + 2) (-5 \cos^2 \! \theta_V + 3)$ &
$\varepsilon \left( |H_+|^2 \! + |H_-|^2 \right)$ \\
$-\frac{20}{3\pi}_{\vphantom\dagger} \sin 2\theta_V \cos \chi$ &
${\rm Re} \left[ \eta (H_+ \! - H_- ) H^*_0  - \varepsilon
( H_+ \! + H_- ) H^*_t \right]$ \\
\hline
\end{tabular}
\end{table*}

\subsection{The Decays $P \! \to \! P' \ell \nu_\ell$}

The much simpler class of decays $P \! \to \! P' \ell \nu_\ell$, where
$P'$ like $P$ is also a pseudoscalar meson, is presented here,
following the more complicated class $P \! \to \! V \ell \nu_\ell$, $V
\! \to \! P_1 P_2$, because the relevant partial-wave expressions can
be deduced almost immediately from the previous case.  One notes that
since the $P'$ is spinless, the $W^*$ can couple only through its
helicity-0 states: the $J \! = \! 1$ component that couples to the
helicity amplitude $H_0$, and the $J \! = \! 0$ component that couples
to the helicity amplitude $H_t$.  To be specific, the form factors for
the transition of a pseudoscalar meson $P$ (mass $M$, momentum $p$) to
a pseudoscalar meson $P'$ (mass $m$, momentum $p'$) are defined
as~\cite{Boyd:1997kz}
\begin{eqnarray}
\left< P'(p') \left| V^\mu \right| P(p) \right> & = & f_+(q^2)
( p + p' )^\mu + f_-(q^2) q^\mu \, . \nonumber \\
\end{eqnarray}
Then the helicity amplitudes are given by~\cite{Korner:1989qb}
\begin{eqnarray}
H_0 & = & \frac{2Mp^{\vphantom\dagger}_V}{\sqrt{q^2}} f_+ \,
, \nonumber \\
H_t & = & \frac{1}{\sqrt{q^2}} f_0 = \frac{1}{\sqrt{q^2}}
\left[ ( M^2 - m^2 ) f_+ + q^2 f_- \right] \, , \nonumber \\
\end{eqnarray}
where the combination $f_0$ is defined in Ref.~\cite{Boyd:1997kz}.
Note particularly that the same names $H_0$, $H_t$ are used here for
the helicity amplitudes of $P \! \to \! P' \ell \nu_\ell$ as for $P
\! \to \! V \ell \nu_\ell$, $V \! \to \! P_1 P_2$, even though they
refer to distinct hadronic quantities in the two cases.  The label $V$
in the momentum $p^{\vphantom\dagger}_V$ defined in Eq.~(\ref{eq:pV})
now refers to $P'$ in this subsection.

The full differential rate for $P \! \to \! P' \ell \nu_\ell$ depends
only upon two variables, namely, $q^2$ and $\theta_\ell$, where
$\theta_\ell$ is defined precisely as in Fig.~\ref{fig1}.  One may
obtain the differential rate simply by taking the expression in
Eq.~(\ref{eq:RBwidth}) and setting $H_+ \! = \! 0$, $H_- \! = \! 0$,
${\cal B} (V \! \to \! P_1 P_2) \! = \! 1$, and integrating over the
full ranges of $d\cos \theta_V$ and $d\chi$.\footnote{Strictly
speaking, in Eq.~(\ref{eq:Wigner}) one replaces $\sqrt{2 \! \cdot \! 1
\! + \! 1} \, d^1_{0,0} (\cos \theta_V) \! = \! \sqrt{3} \cos
\theta_V$ with $\sqrt{2 \! \cdot \! 0 \! + \! 1} \, d^0_{0,0} (\cos
\theta_V) \! = \! 1$; integrating over $d\cos
\theta_V$ in either case then gives $+2$.}  One obtains
\begin{eqnarray}
\lefteqn{\frac{d\Gamma (P \! \to \! P' \ell \nu_\ell)}{dq^2 d \! \cos
\theta_\ell} = 
\frac{1}{128\pi^3} G_F^2 |V_{q'Q}|^2
\frac{p^{\vphantom\dagger}_V q^2 (1 \! - \! \varepsilon)^2}{M^2}}
\nonumber \\ & &\phantom{xxx}\times  \left[ \left( \sin^2 \!
\theta_\ell + \varepsilon \cos^2 \! \theta_\ell \right) \left| H_0
(q^2) \right|^2 \right. \nonumber \\
& &\phantom{xxxxx} \left. - 2\varepsilon \cos \theta_\ell \, {\rm Re}
\, H_0 H_t^* (q^2) + \varepsilon \left| H_t (q^2) \right|^2\right] \, .
\label{eq:BDwidth}
\end{eqnarray}
Clearly, being able to use the same names $H_0$, $H_t$ for both $P
\! \to \! P' \ell \nu_\ell$ and $P \! \to \! V \ell \nu_\ell$, $V \!
\to \! P_1 P_2$ in the reduction of Eq.~(\ref{eq:RBwidth}) means that
the helicity amplitudes must have the correct relative normalization.
One may also integrate over the full range of $\theta_\ell$ to obtain
\begin{eqnarray}
\frac{d\Gamma}{dq^2} & = & \frac{d\Gamma_0}{dq^2} \left\{
\left( 1 + \frac{\varepsilon}{2} \right) |H_0|^2 + \frac 3 2
\varepsilon |H_t|^2 \right\} \, ,
\label{eq:IntWidth2}
\end{eqnarray}
which precisely matches Eq.~(\ref{eq:IntWidth}) after setting $H_+ \!
= \! 0$, $H_- \! = \! 0$ (again indicating the proper relative
normalization between these $H_0$, $H_t$ helicity amplitudes and the
ones of the same names for the case $P \! \to \! V$).  The overall
differential width coefficient in Eq.~(\ref{eq:IntWidth2}),
\begin{equation}
\frac{d\Gamma_0}{dq^2} \equiv \frac{1}{96\pi^3} G_F^2 \left| V_{q' Q}
\right|^2 \frac{p^{\vphantom\dagger}_V q^2 (1-\varepsilon)^2}{M^2} \,
,
\label{eq:overall2}
\end{equation}
assumes the same form as in Eq.~(\ref{eq:overall}), except that now
${\cal B} (V \to P_1 P_2) \! = \! 1$.

The particular weight functions $w_0 (\theta_\ell)$ analogous to those
in Table~\ref{tab:Weight1} are defined as ones that extract simple
helicity amplitude combinations when performing integrals analogous to
those in Eq.~(\ref{eq:Weight}):
\begin{eqnarray} \label{eq:Weight2}
& & \left( \frac{d\Gamma_0}{dq^2} \right)^{\! -1} \! \! \int_{-1}^{+1}
d \! \cos \theta_\ell \, w_0(\theta_\ell) \frac{d\Gamma}
{dq^2\, d \! \cos \theta_\ell } \, . \ \
\end{eqnarray}
The required weight functions $w_0(\theta_\ell)$ and the 3 independent
simple combinations of helicity amplitudes that can be extracted are
listed in Table~\ref{tab:Weight1b}.  One notes that these combinations
are precisely the subset of those in Table~\ref{tab:Weight1} depending
only upon $H_0$ and $H_t$ (although, again, they refer here to $P \!
\to P'$ and not $P \! \to \! V$ transitions).

\begin{table}[ht]
\setlength\extrarowheight{2.5pt}
\caption{Weight functions $w_0(\theta_\ell)$ integrated against the
full 2-fold differential width of Eq.~(\ref{eq:BDwidth}) for processes
$P \! \to P' \ell \nu_\ell$ in the manner described in
Eq.~(\ref{eq:Weight2}).}
\label{tab:Weight1b}
\begin{tabular}{cc}
$w_0(\theta_\ell)$ & {\rm Extracted helicity amplitude}
\\
\hline
$\frac 5 2 \left( -3\cos^2 \! \theta_\ell + 1 \right)$ & $| H_0 |^2
(1 - \varepsilon)$ \\
$-\cos \theta_\ell$ & $\varepsilon \, {\rm Re} \, H_0 H^*_t$ \\
$5 \cos^2 \! \theta_\ell - 1$ & $\varepsilon \left( |H_0|^2 + |H_t|^2
\right)$ \\
\hline
\end{tabular}
\end{table}

\section{$B_c \! \to \! J/\psi \, \ell \nu_\ell$}
\label{sec:Bc}

The corresponding results for $P \! \to \! V \ell^\prime \nu$, $V \!
\to \! \ell^- \ell^+$ can be obtained in an analogous way.  Gone is
the simplification of the previous case, in which the spinless $P_1$
and $P_2$ both have zero helicity.  However, in the physically
relevant case of $B_c \! \to \! J/\psi \, \ell \nu$, $J/\psi \! \to \!
\tilde \ell^- \tilde \ell^+$, the $J/\psi$ is too light to decay to
$\tau^+ \tau^-$, while for $\tilde \ell \! = \! \mu$ (the
experimentally favored channel for reconstruction of a $J/\psi$), one
has $(m_\mu / m_{J/\psi})^2 \! = \! 1.16 \cdot 10^{-3}$: The outgoing
$\mu$ pair are almost pure helicity eigenstates, a restriction that
reduces the angular analysis to be almost as straightforward as in the
previous section.  We thus ignore $m_\mu$ in the decay of $J/\psi$ but
retain $m_\ell$ from the semileptonic decay.

The expansion of Eq.~(\ref{eq:Wigner}) holds for this new case, with
the notational substitution of $P \! \to \! V \ell \nu$, $V \!
\to \! \tilde \ell^- \tilde \ell^+$.  The ``0'' subscript in
Eq.~(\ref{eq:Wigner}) is replaced by $\sigma \! \equiv \! \tilde
\lambda_{\ell^-} \! \!  - \! \tilde \lambda_{\ell^+}$.  One
immediately notes that the two $\sigma \! = \! 0$ cases of $\tilde
\ell^-_L \tilde \ell^+_L$ and $\tilde \ell^-_R \tilde \ell^+_R$ give
results algebraically identical to Eq.~(\ref{eq:RBwidth}),
upon substituting ${\cal B} (V \! \to \! P_1 P_2)$ with ${\cal B} (V
\! \to \! \tilde \ell^- \tilde \ell^+)$, and the results of
Table~\ref{tab:Weight1} apply equally well for the two $\sigma \! = \!
0$ cases.  Note the identification of $P_1 \! \to \! \tilde \ell^-$,
as in Fig.~\ref{fig1}, for the purpose of defining scattering angles.

The opposite-helicity ($\sigma \! = \! \mp 1$) combinations are more
complicated because the rotation matrices on the $V \!
\to \! \tilde \ell^- \tilde \ell^+$ side are nontrivial.  In analogue
to Eq.~(\ref{eq:RBwidth}), and restricting for simplicity to the
case $\nu_\ell \! \to \! \bar \nu_\ell$, one finds
\begin{widetext}
\begin{align}
&\frac{d\Gamma (P \! \to V \! \ell \bar \nu_\ell, \, V \! \to
\tilde \ell^-_L \tilde \ell^+_R)}
{dq^2 \, d \! \cos \theta_V \, d \! \cos \theta_\ell \, d\chi}=
\frac{3G_F^2 \left| V_{q' Q} \right|^2}{8(4 \pi)^4}
\frac{p^{\vphantom\dagger}_V q^2 (1-\varepsilon)^2}{M^2} {\cal B}
(V \to \tilde \ell^-_L \tilde \ell^+_R) \left\{ 8 \sin^2 \!
\frac{\theta_\ell}{2}
\left( \sin^2 \! \frac{\theta_\ell}{2} + \varepsilon \cos^2 \!
\frac{\theta_\ell}{2} \right) \sin^4 \! \frac{\theta_V}{2}
\left| H_+ (q^2) \right|^2 \right. \nonumber \\ 
&\phantom{xxx} +8 \cos^2 \! \frac{\theta_\ell}{2}
\left( \cos^2 \! \frac{\theta_\ell}{2} + \varepsilon \sin^2 \!
\frac{\theta_\ell}{2} \right) \! \cos^4 \! \frac{\theta_V}{2}
\left| H_- (q^2) \right|^2 + \, 2 \left( \sin^2 \! \theta_\ell +
\varepsilon \cos^2 \! \theta_\ell \right) \sin^2 \! \theta_V
\left| H_0 (q^2) \right|^2 \nonumber\\
&\phantom{xxx}- 8 \sin \theta_\ell \sin \theta_V \! \cos \chi 
\left[\left( \sin^2 \! \frac{\theta_\ell}{2} + \frac{\varepsilon}{2}
\cos \theta_\ell \right) \sin^2 \! \frac{\theta_V}{2} \,
{\rm Re} \, H_+ H^*_0 (q^2) + \left( \cos^2 \! \frac{\theta_\ell}{2} -
\frac{\varepsilon}{2}
\cos \theta_\ell \right) \! \cos^2 \! \frac{\theta_V}{2}
{\rm Re} \, H_- H^*_0 (q^2)  \right] \nonumber \\ 
&\phantom{xxx}+ \sin^2 \! \theta_\ell \sin^2 \! \theta_V \cos 2\chi \,
( 1 - \varepsilon ) \, {\rm Re} \, H_+ H^*_- (q^2) + \, 2\varepsilon
\sin^2 \! \theta_V \! \left[
\left| H_t (q^2) \right|^2 \! -2 \cos \theta_\ell \, {\rm Re} \,
H_t H^*_0 (q^2) \right] \nonumber \\ 
&\phantom{xxx} + 4 \varepsilon \sin
\theta_\ell \sin \theta_V \cos \chi \! \left. \left[ \sin^2
\frac{\theta_V}{2} {\rm Re} \, H_+ H^*_t \! (q^2) - \cos^2
\frac{\theta_V}{2} {\rm Re} \, H_- H^*_t \! (q^2)\right] \right\} .
\label{eq:RBwidth2}
\end{align}
\end{widetext}
The corresponding expression for $\bar \nu_\ell \! \to \! \nu_\ell$ is
obtained by exchanging $\sin^2 ( \theta_\ell / 2 ) \leftrightarrow
\cos^2 ( \theta_\ell / 2 )$ throughout Eq.~(\ref{eq:RBwidth2}), with
the insertion of an additional sign on these coefficients in the
${\rm Re} \, H_\pm H^*_0$ terms.\footnote{This result is the effect
of $d^1_{\lambda, +1} (\theta_\ell) \! \to \!  d^1_{\lambda, -1}
(\theta_\ell)$ in the relevant terms, which effectively takes
$\theta_\ell \! \to \! \theta_\ell \! + \! \pi$; had we retained
CP-violating terms, one would find from the phase in the full rotation
matrix that $\sin \chi \! \to \! -\sin \chi$ as well.}  The
corresponding expression for $\tilde \ell^-_L \tilde
\ell^+_R \! \to \tilde \ell^-_R \tilde \ell^+_L$ is obtained by
exchanging $\sin^2 ( \theta_\ell / 2 ) \leftrightarrow \cos^2 (
\theta_\ell / 2 )$, as well as $\sin \theta_V \! \to \! -\sin
\theta_V$, throughout Eq.~(\ref{eq:RBwidth2}).\footnote{This result is
the effect of $d^1_{\lambda, +1} (\theta_V) \! \leftrightarrow \!
d^1_{\lambda, -1} (\theta_V)$ in the relevant terms, which effectively
takes $\theta_V \! \to \! \theta_V \! + \! \pi$.}  One can then derive
simple weight functions analogous to those used in
Table~\ref{tab:Weight1} to obtain the results for $\tilde \ell^-_L
\tilde \ell^+_R$ and $\tilde \ell^-_R \tilde \ell^+_L$ given in
Table~\ref{tab:Weight2}.

\begin{table}[ht]
\setlength\extrarowheight{2.5pt}
\caption{Weight functions $w_\sigma (\theta_\ell, \theta_V \! ,
\chi)$ integrated against the full 4-fold differential width
Eq.~(\ref{eq:RBwidth2}) for processes $P \! \to V \! \ell \bar \nu,
\, V \! \to \tilde \ell^-_L \tilde \ell^+_R$ in the manner described
in Eq.~(\ref{eq:Weight}) (with $w_0 \! \to w_\sigma$).  They apply in
cases where the $V$ decays to (massless) leptons with total
spin-projection $\sigma \! = \! \mp 1$ (which correspond to $\tilde
\ell^-_L \tilde \ell^+_R$ and $\tilde \ell^-_R \tilde \ell^+_L$,
respectively) along the decay axis.}
\label{tab:Weight2}
\begin{tabular}{cc}
$w_\sigma (\theta_\ell, \theta_V \! , \chi)$ &
\shortstack{\rm Extracted\\ helicity amplitude} \\
\hline
$(-5 \cos^2 \! \theta_\ell + 1) (-5 \cos^2 \! \theta_V + 1)$
& $|H_+|^2 \! + |H_-|^2$ \\
$\frac 5 2 (-3 \cos^2 \! \theta_\ell + 1) (-5 \cos^2 \! \theta_V +2)$
& $|H_0|^2 ( 1 - \varepsilon )$ \\
$+2\eta \cos \theta_\ell (-5 \cos^2 \! \theta_V + 1)$ &
$|H_+|^2 \! - |H_-|^2$ \\
$-\frac{20}{3\pi} \sin 2\theta_\ell (\sigma + 4 \cos \theta_V \! )
\cos \chi$ & $( 1 - \varepsilon ) {\rm Re} (H_+ H^*_0)$
\\
$+\frac{20}{3\pi} \sin 2\theta_\ell (\sigma - 4 \cos \theta_V \! )
\cos \chi$ & $( 1 - \varepsilon ) {\rm Re} (H_- H^*_0)$
\\
$4 \cos 2\chi$ & $(1 - \varepsilon) {\rm Re} \, H_+ H^*_-$ \\
$-\cos \theta_\ell (-5 \cos^2 \! \theta_V + 2)$ &
$\varepsilon \, {\rm Re} \, H_0 H^*_t$ \\
$- (-5 \cos^2 \! \theta_\ell + 1) (-5 \cos^2 \! \theta_V + 2)$ &
$\varepsilon \left( |H_0|^2 + |H_t|^2 \right)$ \\
$-2(-5 \! \cos^2 \! \theta_\ell + 2) (-5 \cos^2 \! \theta_V + 1)$ &
$\varepsilon \left( |H_+|^2 \! + |H_-|^2 \right)$ \\
$+4\sigma (-5 \cos^2 \! \theta_\ell + 2) \cos \theta_V$ &
$\varepsilon \left( |H_+|^2 \! - |H_-|^2 \right)$ \\
$+\frac{8}{3\pi} (1 + 5 \eta \cos \theta_\ell) (\sigma + 4 \cos
\theta_V \! ) \cos \chi$ & $\varepsilon {\rm Re} \left[ H_+ \!
\left( \eta H^*_0 - H^*_t \right) \right]$ \\
$+\frac{8}{3\pi}_{\vphantom\dagger} (1 - 5 \eta \cos \theta_\ell)
(\sigma - 4 \cos \theta_V \! ) \cos \chi$ & $\varepsilon {\rm Re}
\left[ H_- \! \left( \eta H^*_0 + H^*_t \right) \right]$ \\
\hline
\end{tabular}
\end{table}

From Table~\ref{tab:Weight2}, one immediately notes that additional
combinations of helicity amplitudes can be extracted from the data
independently of the lepton mass parameter $\varepsilon$.  While
Table~\ref{tab:Weight1} shows that 9 of the 16 possible
combinations\footnote{Again, $H_t$ only appears with coefficient
$\varepsilon$.} ${\rm Re} \, H_i H^*_j$, $\varepsilon {\rm Re} \, H_i
H^*_j$ can be isolated using appropriate weight functions $w_0
(\theta_\ell, \theta_V \! , \chi)$, Table~\ref{tab:Weight1} shows that
12 combinations can be isolated when one has complete polarization
information on the $\tilde \ell^\pm$ pair.  7 of the 12 combinations
in Table~\ref{tab:Weight2} also appear verbatim in
Table~\ref{tab:Weight1}; in addition, the new combination $\varepsilon
\left( |H_+|^2 \! - |H_-|^2 \right)$ appears, and the 2 remaining
combinations in Table~\ref{tab:Weight1} appear as linear combinations
of the 4 entries of Table~\ref{tab:Weight2} with $w_\sigma$
proportional to $\cos \chi$.  That is to say, the entries of
Table~\ref{tab:Weight1} do not provide access to any combinations
independent of those in Table~\ref{tab:Weight2}.

That 4 linear combinations of helicity amplitude combinations remain
inaccessible even in the case in which the polarization state of the
$V$ is well probed via access to the $\tilde \ell^\pm$ helicities once
again points to the restrictiveness of the underlying $V \! - \! A$
interaction.  Nevertheless, the redundancy of some amplitude
combinations provides a precise handle on probing non-SM effects.  For
example, access to the amplitude combination $\varepsilon \left(
|H_+|^2 \! - |H_-|^2 \right)$, in addition to the combination $\left(
|H_+|^2 \! - |H_-|^2 \right)$, provides another very clean
determination of $\varepsilon$, completely analogous to but separate
from that of Eqs.~(\ref{eq:extracte1})--(\ref{eq:extracte2}), or tests
analogous to those in Eqs.~(\ref{eq:rvel})--(\ref{eq:Weight18}).

\section{Conclusions}
\label{sec:Concl}

In this paper we have constructed robust tests of generic
lepton-universality violations in semileptonic decays that are
independent of knowledge of the transition form factors between
hadronic states, particularly for a pseudoscalar meson (such as $B$ or
$B_c$) decaying to a hadron $H$ (such as $D^*$ or $D$ or $J/\psi$).
Starting from the fully differential cross section decomposed into the
helicity basis, one can construct weight functions that project onto
specific combinations, labeled by $i$, of helicity amplitudes.
Integrating the differential cross section in different lepton
channels against these weight functions and taking their ratios
$R^H_i$, the entire form-factor dependence is eliminated, and the
Standard Model predicts unity for these ratios.  We furthermore found
an infinite class of such relations, based upon how one chooses to
weight combinations corresponding to the various amplitudes $i$, and
we also found analogous relations even within processes of a {\em
single\/} lepton flavor.

The occurrence $R^H_i\neq 1$ for some ratio $i$ does not necessarily
imply lepton-universality violation, but it does require BSM of some
form that acts differently for different final-state leptons.  If one
attributes the current tension in the measured ratios $R(H)$ to BSM,
our tests provide a deeper level of information.  Either at least one
of the $R^H_i$ must differ from unity, thereby suggesting the
structure of the BSM physics based upon which helicity combination
exhibits this signal; or else no non-unity $R^H_i$ is found, in which
case the BSM must reside in the $q^2\leq m_\tau^2$ muon data ({\it
i.e.}, the nonuniversal portion of the lepton phase space).  In that
scenario, other muonic tests like Eq.~(\ref{eq:rvel})---a
single-lepton flavor test that uses the entire phase space---or
$(g-2)_\mu$ can provide constraints.

\begin{acknowledgments}
  \vspace{-2ex} This work was supported by the U.S.\ Department of
  Energy under Contract No.\ DE-FG02-93ER-40762 (T.D.C.\ and H.L.) and
  the National Science Foundation under Grant No.\ PHY-1403891
  (R.F.L.).
\end{acknowledgments}

\bibliographystyle{apsrev4-1}
\bibliography{wise}

\begin{thebibliography}{43}%
\makeatletter
\providecommand \@ifxundefined [1]{%
 \@ifx{#1\undefined}
}%
\providecommand \@ifnum [1]{%
 \ifnum #1\expandafter \@firstoftwo
 \else \expandafter \@secondoftwo
 \fi
}%
\providecommand \@ifx [1]{%
 \ifx #1\expandafter \@firstoftwo
 \else \expandafter \@secondoftwo
 \fi
}%
\providecommand \natexlab [1]{#1}%
\providecommand \enquote  [1]{``#1''}%
\providecommand \bibnamefont  [1]{#1}%
\providecommand \bibfnamefont [1]{#1}%
\providecommand \citenamefont [1]{#1}%
\providecommand \href@noop [0]{\@secondoftwo}%
\providecommand \href [0]{\begingroup \@sanitize@url \@href}%
\providecommand \@href[1]{\@@startlink{#1}\@@href}%
\providecommand \@@href[1]{\endgroup#1\@@endlink}%
\providecommand \@sanitize@url [0]{\catcode `\\12\catcode `\$12\catcode
  `\&12\catcode `\#12\catcode `\^12\catcode `\_12\catcode `\%12\relax}%
\providecommand \@@startlink[1]{}%
\providecommand \@@endlink[0]{}%
\providecommand \url  [0]{\begingroup\@sanitize@url \@url }%
\providecommand \@url [1]{\endgroup\@href {#1}{\urlprefix }}%
\providecommand \urlprefix  [0]{URL }%
\providecommand \Eprint [0]{\href }%
\providecommand \doibase [0]{http://dx.doi.org/}%
\providecommand \selectlanguage [0]{\@gobble}%
\providecommand \bibinfo  [0]{\@secondoftwo}%
\providecommand \bibfield  [0]{\@secondoftwo}%
\providecommand \translation [1]{[#1]}%
\providecommand \BibitemOpen [0]{}%
\providecommand \bibitemStop [0]{}%
\providecommand \bibitemNoStop [0]{.\EOS\space}%
\providecommand \EOS [0]{\spacefactor3000\relax}%
\providecommand \BibitemShut  [1]{\csname bibitem#1\endcsname}%
\let\auto@bib@innerbib\@empty
\bibitem [{\citenamefont {Amhis}\ \emph {et~al.}(2017)\citenamefont {Amhis}
  \emph {et~al.}}]{Amhis:2016xyh}%
  \BibitemOpen
  \bibfield  {author} {\bibinfo {author} {\bibfnamefont {Y.}~\bibnamefont
  {Amhis}} \emph {et~al.} (\bibinfo {collaboration} {HFLAV Group}),\ }\href
  {\doibase 10.1140/epjc/s10052-017-5058-4} {\bibfield  {journal} {\bibinfo
  {journal} {Eur.\ Phys.\ J.}\ }\textbf {\bibinfo {volume} {C77}},\ \bibinfo
  {pages} {895} (\bibinfo {year} {2017})},\ \Eprint
  {http://arxiv.org/abs/1612.07233} {arXiv:1612.07233 [hep-ex]} \BibitemShut
  {NoStop}%
\bibitem [{\citenamefont {Lees}\ \emph {et~al.}(2012)\citenamefont {Lees} \emph
  {et~al.}}]{Lees:2012xj}%
  \BibitemOpen
  \bibfield  {author} {\bibinfo {author} {\bibfnamefont {J.~P.}\ \bibnamefont
  {Lees}} \emph {et~al.} (\bibinfo {collaboration} {BaBar Collaboration}),\
  }\href {\doibase 10.1103/PhysRevLett.109.101802} {\bibfield  {journal}
  {\bibinfo  {journal} {Phys.\ Rev.\ Lett.}\ }\textbf {\bibinfo {volume}
  {109}},\ \bibinfo {pages} {101802} (\bibinfo {year} {2012})},\ \Eprint
  {http://arxiv.org/abs/1205.5442} {arXiv:1205.5442 [hep-ex]} \BibitemShut
  {NoStop}%
\bibitem [{\citenamefont {Lees}\ \emph {et~al.}(2013)\citenamefont {Lees} \emph
  {et~al.}}]{Lees:2013uzd}%
  \BibitemOpen
  \bibfield  {author} {\bibinfo {author} {\bibfnamefont {J.~P.}\ \bibnamefont
  {Lees}} \emph {et~al.} (\bibinfo {collaboration} {BaBar Collaboration}),\
  }\href {\doibase 10.1103/PhysRevD.88.072012} {\bibfield  {journal} {\bibinfo
  {journal} {Phys.\ Rev.}\ }\textbf {\bibinfo {volume} {D88}},\ \bibinfo
  {pages} {072012} (\bibinfo {year} {2013})},\ \Eprint
  {http://arxiv.org/abs/1303.0571} {arXiv:1303.0571 [hep-ex]} \BibitemShut
  {NoStop}%
\bibitem [{\citenamefont {Huschle}\ \emph {et~al.}(2015)\citenamefont {Huschle}
  \emph {et~al.}}]{Huschle:2015rga}%
  \BibitemOpen
  \bibfield  {author} {\bibinfo {author} {\bibfnamefont {M.}~\bibnamefont
  {Huschle}} \emph {et~al.} (\bibinfo {collaboration} {Belle Collaboration}),\
  }\href {\doibase 10.1103/PhysRevD.92.072014} {\bibfield  {journal} {\bibinfo
  {journal} {Phys.\ Rev.}\ }\textbf {\bibinfo {volume} {D92}},\ \bibinfo
  {pages} {072014} (\bibinfo {year} {2015})},\ \Eprint
  {http://arxiv.org/abs/1507.03233} {arXiv:1507.03233 [hep-ex]} \BibitemShut
  {NoStop}%
\bibitem [{\citenamefont {Sato}\ \emph {et~al.}(2016)\citenamefont {Sato} \emph
  {et~al.}}]{Sato:2016svk}%
  \BibitemOpen
  \bibfield  {author} {\bibinfo {author} {\bibfnamefont {Y.}~\bibnamefont
  {Sato}} \emph {et~al.} (\bibinfo {collaboration} {Belle Collaboration}),\
  }\href {\doibase 10.1103/PhysRevD.94.072007} {\bibfield  {journal} {\bibinfo
  {journal} {Phys.\ Rev.}\ }\textbf {\bibinfo {volume} {D94}},\ \bibinfo
  {pages} {072007} (\bibinfo {year} {2016})},\ \Eprint
  {http://arxiv.org/abs/1607.07923} {arXiv:1607.07923 [hep-ex]} \BibitemShut
  {NoStop}%
\bibitem [{\citenamefont {Aaij}\ \emph {et~al.}(2015)\citenamefont {Aaij} \emph
  {et~al.}}]{Aaij:2015yra}%
  \BibitemOpen
  \bibfield  {author} {\bibinfo {author} {\bibfnamefont {R.}~\bibnamefont
  {Aaij}} \emph {et~al.} (\bibinfo {collaboration} {LHCb Collaboration}),\
  }\href {\doibase 10.1103/PhysRevLett.115.159901,
  10.1103/PhysRevLett.115.111803} {\bibfield  {journal} {\bibinfo  {journal}
  {Phys.\ Rev.\ Lett.}\ }\textbf {\bibinfo {volume} {115}},\ \bibinfo {pages}
  {111803} (\bibinfo {year} {2015})},\ \bibinfo {note} {[Erratum: Phys.\ Rev.\
  Lett.\ {\bf 115}, 159901 (2015)]},\ \Eprint {http://arxiv.org/abs/1506.08614}
  {arXiv:1506.08614 [hep-ex]} \BibitemShut {NoStop}%
\bibitem [{\citenamefont {Hirose}\ \emph {et~al.}(2017)\citenamefont {Hirose}
  \emph {et~al.}}]{Hirose:2016wfn}%
  \BibitemOpen
  \bibfield  {author} {\bibinfo {author} {\bibfnamefont {S.}~\bibnamefont
  {Hirose}} \emph {et~al.} (\bibinfo {collaboration} {Belle Collaboration}),\
  }\href {\doibase 10.1103/PhysRevLett.118.211801} {\bibfield  {journal}
  {\bibinfo  {journal} {Phys.\ Rev.\ Lett.}\ }\textbf {\bibinfo {volume}
  {118}},\ \bibinfo {pages} {211801} (\bibinfo {year} {2017})},\ \Eprint
  {http://arxiv.org/abs/1612.00529} {arXiv:1612.00529 [hep-ex]} \BibitemShut
  {NoStop}%
\bibitem [{\citenamefont {Wormser}(2017)}]{Wormser:2017hsx}%
  \BibitemOpen
  \bibfield  {author} {\bibinfo {author} {\bibfnamefont {G.}~\bibnamefont
  {Wormser}} (\bibinfo {collaboration} {BaBar, LHCb Collaborations}),\
  }\bibfield  {booktitle} {\emph {\bibinfo {booktitle} {{Proceedings, 15th
  Conference on Flavor Physics and CP Violation (FPCP 2017): Prague, Czech
  Republic, June 5--9, 2017}}},\ }\href {\doibase 10.22323/1.304.0006}
  {\bibfield  {journal} {\bibinfo  {journal} {PoS FPCP}\ }\textbf {\bibinfo
  {volume} {2017}},\ \bibinfo {pages} {006} (\bibinfo {year}
  {2017})}\BibitemShut {NoStop}%
\bibitem [{\citenamefont {Aaij}\ \emph {et~al.}(2018)\citenamefont {Aaij} \emph
  {et~al.}}]{Aaij:2017tyk}%
  \BibitemOpen
  \bibfield  {author} {\bibinfo {author} {\bibfnamefont {R.}~\bibnamefont
  {Aaij}} \emph {et~al.} (\bibinfo {collaboration} {LHCb Collaboration}),\
  }\href {\doibase 10.1103/PhysRevLett.120.121801} {\bibfield  {journal}
  {\bibinfo  {journal} {Phys.\ Rev.\ Lett.}\ }\textbf {\bibinfo {volume}
  {120}},\ \bibinfo {pages} {121801} (\bibinfo {year} {2018})},\ \Eprint
  {http://arxiv.org/abs/1711.05623} {arXiv:1711.05623 [hep-ex]} \BibitemShut
  {NoStop}%
\bibitem [{\citenamefont {Aoki}\ \emph {et~al.}(2017)\citenamefont {Aoki} \emph
  {et~al.}}]{Aoki:2016frl}%
  \BibitemOpen
  \bibfield  {author} {\bibinfo {author} {\bibfnamefont {S.}~\bibnamefont
  {Aoki}} \emph {et~al.},\ }\href {\doibase 10.1140/epjc/s10052-016-4509-7}
  {\bibfield  {journal} {\bibinfo  {journal} {Eur.\ Phys.\ J.}\ }\textbf
  {\bibinfo {volume} {C77}},\ \bibinfo {pages} {112} (\bibinfo {year}
  {2017})},\ \Eprint {http://arxiv.org/abs/1607.00299} {arXiv:1607.00299
  [hep-lat]} \BibitemShut {NoStop}%
\bibitem [{\citenamefont {Bailey}\ \emph {et~al.}(2015)\citenamefont {Bailey}
  \emph {et~al.}}]{Lattice:2015rga}%
  \BibitemOpen
  \bibfield  {author} {\bibinfo {author} {\bibfnamefont {J.~A.}\ \bibnamefont
  {Bailey}} \emph {et~al.} (\bibinfo {collaboration} {MILC Collaboration}),\
  }\href {\doibase 10.1103/PhysRevD.92.034506} {\bibfield  {journal} {\bibinfo
  {journal} {Phys.\ Rev.}\ }\textbf {\bibinfo {volume} {D92}},\ \bibinfo
  {pages} {034506} (\bibinfo {year} {2015})},\ \Eprint
  {http://arxiv.org/abs/1503.07237} {arXiv:1503.07237 [hep-lat]} \BibitemShut
  {NoStop}%
\bibitem [{\citenamefont {Na}\ \emph {et~al.}(2015)\citenamefont {Na},
  \citenamefont {Bouchard}, \citenamefont {Lepage}, \citenamefont {Monahan},\
  and\ \citenamefont {Shigemitsu}}]{Na:2015kha}%
  \BibitemOpen
  \bibfield  {author} {\bibinfo {author} {\bibfnamefont {H.}~\bibnamefont
  {Na}}, \bibinfo {author} {\bibfnamefont {C.~M.}\ \bibnamefont {Bouchard}},
  \bibinfo {author} {\bibfnamefont {G.~P.}\ \bibnamefont {Lepage}}, \bibinfo
  {author} {\bibfnamefont {C.}~\bibnamefont {Monahan}}, \ and\ \bibinfo
  {author} {\bibfnamefont {J.}~\bibnamefont {Shigemitsu}} (\bibinfo
  {collaboration} {HPQCD}),\ }\href {\doibase 10.1103/PhysRevD.93.119906,
  10.1103/PhysRevD.92.054510} {\bibfield  {journal} {\bibinfo  {journal}
  {Phys.\ Rev.}\ }\textbf {\bibinfo {volume} {D92}},\ \bibinfo {pages} {054510}
  (\bibinfo {year} {2015})},\ \bibinfo {note} {[Erratum: Phys.\ Rev.\ {\bf
  D93}, 119906 (2016)]},\ \Eprint {http://arxiv.org/abs/1505.03925}
  {arXiv:1505.03925 [hep-lat]} \BibitemShut {NoStop}%
\bibitem [{\citenamefont {Bigi}\ and\ \citenamefont
  {Gambino}(2016)}]{Bigi:2016mdz}%
  \BibitemOpen
  \bibfield  {author} {\bibinfo {author} {\bibfnamefont {D.}~\bibnamefont
  {Bigi}}\ and\ \bibinfo {author} {\bibfnamefont {P.}~\bibnamefont {Gambino}},\
  }\href {\doibase 10.1103/PhysRevD.94.094008} {\bibfield  {journal} {\bibinfo
  {journal} {Phys.\ Rev.}\ }\textbf {\bibinfo {volume} {D94}},\ \bibinfo
  {pages} {094008} (\bibinfo {year} {2016})},\ \Eprint
  {http://arxiv.org/abs/1606.08030} {arXiv:1606.08030 [hep-ph]} \BibitemShut
  {NoStop}%
\bibitem [{\citenamefont {Dungel}\ \emph {et~al.}(2010)\citenamefont {Dungel}
  \emph {et~al.}}]{Dungel:2010uk}%
  \BibitemOpen
  \bibfield  {author} {\bibinfo {author} {\bibfnamefont {W.}~\bibnamefont
  {Dungel}} \emph {et~al.} (\bibinfo {collaboration} {Belle Collaboration}),\
  }\href {\doibase 10.1103/PhysRevD.82.112007} {\bibfield  {journal} {\bibinfo
  {journal} {Phys.\ Rev.}\ }\textbf {\bibinfo {volume} {D82}},\ \bibinfo
  {pages} {112007} (\bibinfo {year} {2010})},\ \Eprint
  {http://arxiv.org/abs/1010.5620} {arXiv:1010.5620 [hep-ex]} \BibitemShut
  {NoStop}%
\bibitem [{\citenamefont {Fajfer}\ \emph {et~al.}(2012)\citenamefont {Fajfer},
  \citenamefont {Kamenik},\ and\ \citenamefont {Nisandzic}}]{Fajfer:2012vx}%
  \BibitemOpen
  \bibfield  {author} {\bibinfo {author} {\bibfnamefont {S.}~\bibnamefont
  {Fajfer}}, \bibinfo {author} {\bibfnamefont {J.~F.}\ \bibnamefont {Kamenik}},
  \ and\ \bibinfo {author} {\bibfnamefont {I.}~\bibnamefont {Nisandzic}},\
  }\href {\doibase 10.1103/PhysRevD.85.094025} {\bibfield  {journal} {\bibinfo
  {journal} {Phys.\ Rev.}\ }\textbf {\bibinfo {volume} {D85}},\ \bibinfo
  {pages} {094025} (\bibinfo {year} {2012})},\ \Eprint
  {http://arxiv.org/abs/1203.2654} {arXiv:1203.2654 [hep-ph]} \BibitemShut
  {NoStop}%
\bibitem [{\citenamefont {Colquhoun}\ \emph {et~al.}(2016)\citenamefont
  {Colquhoun}, \citenamefont {Davies}, \citenamefont {Koponen}, \citenamefont
  {Lytle},\ and\ \citenamefont {McNeile}}]{Colquhoun:2016osw}%
  \BibitemOpen
  \bibfield  {author} {\bibinfo {author} {\bibfnamefont {B.}~\bibnamefont
  {Colquhoun}}, \bibinfo {author} {\bibfnamefont {C.}~\bibnamefont {Davies}},
  \bibinfo {author} {\bibfnamefont {J.}~\bibnamefont {Koponen}}, \bibinfo
  {author} {\bibfnamefont {A.}~\bibnamefont {Lytle}}, \ and\ \bibinfo {author}
  {\bibfnamefont {C.}~\bibnamefont {McNeile}} (\bibinfo {collaboration}
  {HPQCD}),\ }\bibfield  {booktitle} {\emph {\bibinfo {booktitle}
  {{Proceedings, 34th International Symposium on Lattice Field Theory (Lattice
  2016): Southampton, UK, July 24--30, 2016}}},\ }\href@noop {} {\bibfield
  {journal} {\bibinfo  {journal} {PoS}\ }\textbf {\bibinfo {volume} {LATTICE
  2016}},\ \bibinfo {pages} {281} (\bibinfo {year} {2016})},\ \Eprint
  {http://arxiv.org/abs/1611.01987} {arXiv:1611.01987 [hep-lat]} \BibitemShut
  {NoStop}%
\bibitem [{\citenamefont {Lytle}()}]{ALE}%
  \BibitemOpen
  \bibfield  {author} {\bibinfo {author} {\bibfnamefont {A.}~\bibnamefont
  {Lytle}},\ }\href@noop {} {}\bibinfo {howpublished} {personal
  communication}\BibitemShut {NoStop}%
\bibitem [{\citenamefont {Cohen}\ \emph {et~al.}()\citenamefont {Cohen},
  \citenamefont {Lamm},\ and\ \citenamefont {Lebed}}]{THRprep}%
  \BibitemOpen
  \bibfield  {author} {\bibinfo {author} {\bibfnamefont {T.~D.}\ \bibnamefont
  {Cohen}}, \bibinfo {author} {\bibfnamefont {H.}~\bibnamefont {Lamm}}, \ and\
  \bibinfo {author} {\bibfnamefont {R.~F.}\ \bibnamefont {Lebed}},\ }\href@noop
  {} {\ }\Eprint {http://arxiv.org/abs/1807.02730} {arXiv:1807.02730 [hep-ph]}
  \BibitemShut {NoStop}%
\bibitem [{\citenamefont {Hirose}\ \emph {et~al.}(2018)\citenamefont {Hirose}
  \emph {et~al.}}]{Hirose:2017dxl}%
  \BibitemOpen
  \bibfield  {author} {\bibinfo {author} {\bibfnamefont {S.}~\bibnamefont
  {Hirose}} \emph {et~al.} (\bibinfo {collaboration} {Belle Collaboration}),\
  }\href {\doibase 10.1103/PhysRevD.97.012004} {\bibfield  {journal} {\bibinfo
  {journal} {Phys.\ Rev.}\ }\textbf {\bibinfo {volume} {D97}},\ \bibinfo
  {pages} {012004} (\bibinfo {year} {2018})},\ \Eprint
  {http://arxiv.org/abs/1709.00129} {arXiv:1709.00129 [hep-ex]} \BibitemShut
  {NoStop}%
\bibitem [{\citenamefont {Bailey}\ \emph {et~al.}(2014)\citenamefont {Bailey}
  \emph {et~al.}}]{Bailey:2014tva}%
  \BibitemOpen
  \bibfield  {author} {\bibinfo {author} {\bibfnamefont {J.~A.}\ \bibnamefont
  {Bailey}} \emph {et~al.} (\bibinfo {collaboration} {Fermilab Lattice and MILC
  Collaborations}),\ }\href {\doibase 10.1103/PhysRevD.89.114504} {\bibfield
  {journal} {\bibinfo  {journal} {Phys.\ Rev.}\ }\textbf {\bibinfo {volume}
  {D89}},\ \bibinfo {pages} {114504} (\bibinfo {year} {2014})},\ \Eprint
  {http://arxiv.org/abs/1403.0635} {arXiv:1403.0635 [hep-lat]} \BibitemShut
  {NoStop}%
\bibitem [{\citenamefont {Harrison}\ \emph {et~al.}(2017)\citenamefont
  {Harrison}, \citenamefont {Davies},\ and\ \citenamefont
  {Wingate}}]{Harrison:2016gup}%
  \BibitemOpen
  \bibfield  {author} {\bibinfo {author} {\bibfnamefont {J.}~\bibnamefont
  {Harrison}}, \bibinfo {author} {\bibfnamefont {C.}~\bibnamefont {Davies}}, \
  and\ \bibinfo {author} {\bibfnamefont {M.}~\bibnamefont {Wingate}},\
  }\bibfield  {booktitle} {\emph {\bibinfo {booktitle} {{Proceedings, 34th
  International Symposium on Lattice Field Theory (Lattice 2016): Southampton,
  UK, July 24--30, 2016}}},\ }\href@noop {} {\bibfield  {journal} {\bibinfo
  {journal} {PoS}\ }\textbf {\bibinfo {volume} {LATTICE 2016}},\ \bibinfo
  {pages} {287} (\bibinfo {year} {2017})},\ \Eprint
  {http://arxiv.org/abs/1612.06716} {arXiv:1612.06716 [hep-lat]} \BibitemShut
  {NoStop}%
\bibitem [{\citenamefont {Vaquero Avil{\' e}s-Casco}\ \emph
  {et~al.}(2018)\citenamefont {Vaquero Avil{\' e}s-Casco}, \citenamefont
  {DeTar}, \citenamefont {Du}, \citenamefont {El-Khadra}, \citenamefont
  {Kronfeld}, \citenamefont {Laiho},\ and\ \citenamefont {Van~de
  Water}}]{Aviles-Casco:2017nge}%
  \BibitemOpen
  \bibfield  {author} {\bibinfo {author} {\bibfnamefont {A.}~\bibnamefont
  {Vaquero Avil{\' e}s-Casco}}, \bibinfo {author} {\bibfnamefont
  {C.}~\bibnamefont {DeTar}}, \bibinfo {author} {\bibfnamefont
  {D.}~\bibnamefont {Du}}, \bibinfo {author} {\bibfnamefont {A.}~\bibnamefont
  {El-Khadra}}, \bibinfo {author} {\bibfnamefont {A.~S.}\ \bibnamefont
  {Kronfeld}}, \bibinfo {author} {\bibfnamefont {J.}~\bibnamefont {Laiho}}, \
  and\ \bibinfo {author} {\bibfnamefont {R.~S.}\ \bibnamefont {Van~de Water}},\
  }\bibfield  {booktitle} {\emph {\bibinfo {booktitle} {{Proceedings, 35th
  International Symposium on Lattice Field Theory (Lattice 2017): Granada,
  Spain, June 18-24, 2017}}},\ }\href {\doibase 10.1051/epjconf/201817513003}
  {\bibfield  {journal} {\bibinfo  {journal} {EPJ Web Conf.}\ }\textbf
  {\bibinfo {volume} {175}},\ \bibinfo {pages} {13003} (\bibinfo {year}
  {2018})},\ \Eprint {http://arxiv.org/abs/1710.09817} {arXiv:1710.09817
  [hep-lat]} \BibitemShut {NoStop}%
\bibitem [{\citenamefont {Harrison}\ \emph {et~al.}(2018)\citenamefont
  {Harrison}, \citenamefont {Davies},\ and\ \citenamefont
  {Wingate}}]{Harrison:2017fmw}%
  \BibitemOpen
  \bibfield  {author} {\bibinfo {author} {\bibfnamefont {J.}~\bibnamefont
  {Harrison}}, \bibinfo {author} {\bibfnamefont {C.}~\bibnamefont {Davies}}, \
  and\ \bibinfo {author} {\bibfnamefont {M.}~\bibnamefont {Wingate}} (\bibinfo
  {collaboration} {HPQCD Collaboration}),\ }\href {\doibase
  10.1103/PhysRevD.97.054502} {\bibfield  {journal} {\bibinfo  {journal}
  {Phys.\ Rev.}\ }\textbf {\bibinfo {volume} {D97}},\ \bibinfo {pages} {054502}
  (\bibinfo {year} {2018})},\ \Eprint {http://arxiv.org/abs/1711.11013}
  {arXiv:1711.11013 [hep-lat]} \BibitemShut {NoStop}%
\bibitem [{\citenamefont {Bailey}\ \emph {et~al.}(2018)\citenamefont {Bailey},
  \citenamefont {Bhattacharya}, \citenamefont {Gupta}, \citenamefont {Jang},
  \citenamefont {Lee}, \citenamefont {Leem}, \citenamefont {Park},\ and\
  \citenamefont {Yoon}}]{Bailey:2017xjk}%
  \BibitemOpen
  \bibfield  {author} {\bibinfo {author} {\bibfnamefont {J.~A.}\ \bibnamefont
  {Bailey}}, \bibinfo {author} {\bibfnamefont {T.}~\bibnamefont
  {Bhattacharya}}, \bibinfo {author} {\bibfnamefont {R.}~\bibnamefont {Gupta}},
  \bibinfo {author} {\bibfnamefont {Y.-C.}\ \bibnamefont {Jang}}, \bibinfo
  {author} {\bibfnamefont {W.}~\bibnamefont {Lee}}, \bibinfo {author}
  {\bibfnamefont {J.}~\bibnamefont {Leem}}, \bibinfo {author} {\bibfnamefont
  {S.}~\bibnamefont {Park}}, \ and\ \bibinfo {author} {\bibfnamefont
  {B.}~\bibnamefont {Yoon}} (\bibinfo {collaboration} {LANL-SWME
  Collaboration}),\ }\bibfield  {booktitle} {\emph {\bibinfo {booktitle}
  {{Proceedings, 35th International Symposium on Lattice Field Theory (Lattice
  2017): Granada, Spain, June 18--24, 2017}}},\ }\href {\doibase
  10.1051/epjconf/201817513012} {\bibfield  {journal} {\bibinfo  {journal} {EPJ
  Web Conf.}\ }\textbf {\bibinfo {volume} {175}},\ \bibinfo {pages} {13012}
  (\bibinfo {year} {2018})},\ \Eprint {http://arxiv.org/abs/1711.01786}
  {arXiv:1711.01786 [hep-lat]} \BibitemShut {NoStop}%
\bibitem [{\citenamefont {Bigi}\ \emph {et~al.}(2017)\citenamefont {Bigi},
  \citenamefont {Gambino},\ and\ \citenamefont {Schacht}}]{Bigi:2017jbd}%
  \BibitemOpen
  \bibfield  {author} {\bibinfo {author} {\bibfnamefont {D.}~\bibnamefont
  {Bigi}}, \bibinfo {author} {\bibfnamefont {P.}~\bibnamefont {Gambino}}, \
  and\ \bibinfo {author} {\bibfnamefont {S.}~\bibnamefont {Schacht}},\ }\href
  {\doibase 10.1007/JHEP11(2017)061} {\bibfield  {journal} {\bibinfo  {journal}
  {JHEP}\ }\textbf {\bibinfo {volume} {11}},\ \bibinfo {pages} {061} (\bibinfo
  {year} {2017})},\ \Eprint {http://arxiv.org/abs/1707.09509} {arXiv:1707.09509
  [hep-ph]} \BibitemShut {NoStop}%
\bibitem [{\citenamefont {Colangelo}\ and\ \citenamefont
  {De~Fazio}(2018)}]{Colangelo:2018cnj}%
  \BibitemOpen
  \bibfield  {author} {\bibinfo {author} {\bibfnamefont {P.}~\bibnamefont
  {Colangelo}}\ and\ \bibinfo {author} {\bibfnamefont {F.}~\bibnamefont
  {De~Fazio}},\ }\href {\doibase 10.1007/JHEP06(2018)082} {\bibfield  {journal}
  {\bibinfo  {journal} {JHEP}\ }\textbf {\bibinfo {volume} {06}},\ \bibinfo
  {pages} {082} (\bibinfo {year} {2018})},\ \Eprint
  {http://arxiv.org/abs/1801.10468} {arXiv:1801.10468 [hep-ph]} \BibitemShut
  {NoStop}%
\bibitem [{\citenamefont {Ivanov}\ \emph {et~al.}(2016)\citenamefont {Ivanov},
  \citenamefont {K{\" o}rner},\ and\ \citenamefont {Tran}}]{Ivanov:2016qtw}%
  \BibitemOpen
  \bibfield  {author} {\bibinfo {author} {\bibfnamefont {M.~A.}\ \bibnamefont
  {Ivanov}}, \bibinfo {author} {\bibfnamefont {J.~G.}\ \bibnamefont {K{\"
  o}rner}}, \ and\ \bibinfo {author} {\bibfnamefont {C.-T.}\ \bibnamefont
  {Tran}},\ }\href {\doibase 10.1103/PhysRevD.94.094028} {\bibfield  {journal}
  {\bibinfo  {journal} {Phys.\ Rev.}\ }\textbf {\bibinfo {volume} {D94}},\
  \bibinfo {pages} {094028} (\bibinfo {year} {2016})},\ \Eprint
  {http://arxiv.org/abs/1607.02932} {arXiv:1607.02932 [hep-ph]} \BibitemShut
  {NoStop}%
\bibitem [{\citenamefont {Tran}\ \emph {et~al.}(2018)\citenamefont {Tran},
  \citenamefont {Ivanov}, \citenamefont {K{\" o}rner},\ and\ \citenamefont
  {Santorelli}}]{Tran:2018kuv}%
  \BibitemOpen
  \bibfield  {author} {\bibinfo {author} {\bibfnamefont {C.-T.}\ \bibnamefont
  {Tran}}, \bibinfo {author} {\bibfnamefont {M.~A.}\ \bibnamefont {Ivanov}},
  \bibinfo {author} {\bibfnamefont {J.~G.}\ \bibnamefont {K{\" o}rner}}, \ and\
  \bibinfo {author} {\bibfnamefont {P.}~\bibnamefont {Santorelli}},\ }\href
  {\doibase 10.1103/PhysRevD.97.054014} {\bibfield  {journal} {\bibinfo
  {journal} {Phys.\ Rev.}\ }\textbf {\bibinfo {volume} {D97}},\ \bibinfo
  {pages} {054014} (\bibinfo {year} {2018})},\ \Eprint
  {http://arxiv.org/abs/1801.06927} {arXiv:1801.06927 [hep-ph]} \BibitemShut
  {NoStop}%
\bibitem [{\citenamefont {Bhattacharya}\ \emph {et~al.}(2016)\citenamefont
  {Bhattacharya}, \citenamefont {Nandi},\ and\ \citenamefont
  {Patra}}]{Bhattacharya:2015ida}%
  \BibitemOpen
  \bibfield  {author} {\bibinfo {author} {\bibfnamefont {S.}~\bibnamefont
  {Bhattacharya}}, \bibinfo {author} {\bibfnamefont {S.}~\bibnamefont {Nandi}},
  \ and\ \bibinfo {author} {\bibfnamefont {S.~K.}\ \bibnamefont {Patra}},\
  }\href {\doibase 10.1103/PhysRevD.93.034011} {\bibfield  {journal} {\bibinfo
  {journal} {Phys.\ Rev.}\ }\textbf {\bibinfo {volume} {D93}},\ \bibinfo
  {pages} {034011} (\bibinfo {year} {2016})},\ \Eprint
  {http://arxiv.org/abs/1509.07259} {arXiv:1509.07259 [hep-ph]} \BibitemShut
  {NoStop}%
\bibitem [{\citenamefont {Bhattacharya}\ \emph {et~al.}(2017)\citenamefont
  {Bhattacharya}, \citenamefont {Nandi},\ and\ \citenamefont
  {Patra}}]{Bhattacharya:2016zcw}%
  \BibitemOpen
  \bibfield  {author} {\bibinfo {author} {\bibfnamefont {S.}~\bibnamefont
  {Bhattacharya}}, \bibinfo {author} {\bibfnamefont {S.}~\bibnamefont {Nandi}},
  \ and\ \bibinfo {author} {\bibfnamefont {S.~K.}\ \bibnamefont {Patra}},\
  }\href {\doibase 10.1103/PhysRevD.95.075012} {\bibfield  {journal} {\bibinfo
  {journal} {Phys.\ Rev.}\ }\textbf {\bibinfo {volume} {D95}},\ \bibinfo
  {pages} {075012} (\bibinfo {year} {2017})},\ \Eprint
  {http://arxiv.org/abs/1611.04605} {arXiv:1611.04605 [hep-ph]} \BibitemShut
  {NoStop}%
\bibitem [{\citenamefont {Bhattacharya}\ \emph {et~al.}()\citenamefont
  {Bhattacharya}, \citenamefont {Nandi},\ and\ \citenamefont
  {Kumar~Patra}}]{Bhattacharya:2018kig}%
  \BibitemOpen
  \bibfield  {author} {\bibinfo {author} {\bibfnamefont {S.}~\bibnamefont
  {Bhattacharya}}, \bibinfo {author} {\bibfnamefont {S.}~\bibnamefont {Nandi}},
  \ and\ \bibinfo {author} {\bibfnamefont {S.}~\bibnamefont {Kumar~Patra}},\
  }\href@noop {} {\ }\Eprint {http://arxiv.org/abs/1805.08222}
  {arXiv:1805.08222 [hep-ph]} \BibitemShut {NoStop}%
\bibitem [{\citenamefont {Jaiswal}\ \emph {et~al.}(2017)\citenamefont
  {Jaiswal}, \citenamefont {Nandi},\ and\ \citenamefont
  {Patra}}]{Jaiswal:2017rve}%
  \BibitemOpen
  \bibfield  {author} {\bibinfo {author} {\bibfnamefont {S.}~\bibnamefont
  {Jaiswal}}, \bibinfo {author} {\bibfnamefont {S.}~\bibnamefont {Nandi}}, \
  and\ \bibinfo {author} {\bibfnamefont {S.~K.}\ \bibnamefont {Patra}},\ }\href
  {\doibase 10.1007/JHEP12(2017)060} {\bibfield  {journal} {\bibinfo  {journal}
  {JHEP}\ }\textbf {\bibinfo {volume} {12}},\ \bibinfo {pages} {060} (\bibinfo
  {year} {2017})},\ \Eprint {http://arxiv.org/abs/1707.09977} {arXiv:1707.09977
  [hep-ph]} \BibitemShut {NoStop}%
\bibitem [{\citenamefont {Richman}\ and\ \citenamefont
  {Burchat}(1995)}]{Richman:1995wm}%
  \BibitemOpen
  \bibfield  {author} {\bibinfo {author} {\bibfnamefont {J.~D.}\ \bibnamefont
  {Richman}}\ and\ \bibinfo {author} {\bibfnamefont {P.~R.}\ \bibnamefont
  {Burchat}},\ }\href {\doibase 10.1103/RevModPhys.67.893} {\bibfield
  {journal} {\bibinfo  {journal} {Rev.\ Mod.\ Phys.}\ }\textbf {\bibinfo
  {volume} {67}},\ \bibinfo {pages} {893} (\bibinfo {year} {1995})},\ \Eprint
  {http://arxiv.org/abs/hep-ph/9508250} {arXiv:hep-ph/9508250 [hep-ph]}
  \BibitemShut {NoStop}%
\bibitem [{\citenamefont {Alonso}\ \emph {et~al.}(2017)\citenamefont {Alonso},
  \citenamefont {Martin~Camalich},\ and\ \citenamefont
  {Westhoff}}]{Alonso:2017ktd}%
  \BibitemOpen
  \bibfield  {author} {\bibinfo {author} {\bibfnamefont {R.}~\bibnamefont
  {Alonso}}, \bibinfo {author} {\bibfnamefont {J.}~\bibnamefont
  {Martin~Camalich}}, \ and\ \bibinfo {author} {\bibfnamefont {S.}~\bibnamefont
  {Westhoff}},\ }\href {\doibase 10.1103/PhysRevD.95.093006} {\bibfield
  {journal} {\bibinfo  {journal} {Phys.\ Rev.}\ }\textbf {\bibinfo {volume}
  {D95}},\ \bibinfo {pages} {093006} (\bibinfo {year} {2017})},\ \Eprint
  {http://arxiv.org/abs/1702.02773} {arXiv:1702.02773 [hep-ph]} \BibitemShut
  {NoStop}%
\bibitem [{\citenamefont {Boyd}\ \emph {et~al.}(1997)\citenamefont {Boyd},
  \citenamefont {Grinstein},\ and\ \citenamefont {Lebed}}]{Boyd:1997kz}%
  \BibitemOpen
  \bibfield  {author} {\bibinfo {author} {\bibfnamefont {C.~G.}\ \bibnamefont
  {Boyd}}, \bibinfo {author} {\bibfnamefont {B.}~\bibnamefont {Grinstein}}, \
  and\ \bibinfo {author} {\bibfnamefont {R.~F.}\ \bibnamefont {Lebed}},\ }\href
  {\doibase 10.1103/PhysRevD.56.6895} {\bibfield  {journal} {\bibinfo
  {journal} {Phys.\ Rev.}\ }\textbf {\bibinfo {volume} {D56}},\ \bibinfo
  {pages} {6895} (\bibinfo {year} {1997})},\ \Eprint
  {http://arxiv.org/abs/hep-ph/9705252} {arXiv:hep-ph/9705252 [hep-ph]}
  \BibitemShut {NoStop}%
\bibitem [{\citenamefont {K{\" o}rner}\ and\ \citenamefont
  {Schuler}(1990)}]{Korner:1989qb}%
  \BibitemOpen
  \bibfield  {author} {\bibinfo {author} {\bibfnamefont {J.~G.}\ \bibnamefont
  {K{\" o}rner}}\ and\ \bibinfo {author} {\bibfnamefont {G.~A.}\ \bibnamefont
  {Schuler}},\ }\href {\doibase 10.1007/BF02440838} {\bibfield  {journal}
  {\bibinfo  {journal} {Z. Phys.}\ }\textbf {\bibinfo {volume} {C46}},\
  \bibinfo {pages} {93} (\bibinfo {year} {1990})}\BibitemShut {NoStop}%
\bibitem [{\citenamefont {Gilman}\ and\ \citenamefont
  {Singleton}(1990)}]{Gilman:1989uy}%
  \BibitemOpen
  \bibfield  {author} {\bibinfo {author} {\bibfnamefont {F.~J.}\ \bibnamefont
  {Gilman}}\ and\ \bibinfo {author} {\bibfnamefont {R.~L.}\ \bibnamefont
  {Singleton}},\ }\href {\doibase 10.1103/PhysRevD.41.142} {\bibfield
  {journal} {\bibinfo  {journal} {Phys.\ Rev.}\ }\textbf {\bibinfo {volume}
  {D41}},\ \bibinfo {pages} {142} (\bibinfo {year} {1990})}\BibitemShut
  {NoStop}%
\bibitem [{\citenamefont {Abdesselam}\ \emph {et~al.}()\citenamefont
  {Abdesselam} \emph {et~al.}}]{Abdesselam:2017kjf}%
  \BibitemOpen
  \bibfield  {author} {\bibinfo {author} {\bibfnamefont {A.}~\bibnamefont
  {Abdesselam}} \emph {et~al.} (\bibinfo {collaboration} {Belle
  Collaboration}),\ }\href@noop {} {\ }\Eprint
  {http://arxiv.org/abs/1702.01521} {arXiv:1702.01521 [hep-ex]} \BibitemShut
  {NoStop}%
\bibitem [{\citenamefont {Ji}\ and\ \citenamefont {Lebed}(2001)}]{Ji:2000id}%
  \BibitemOpen
  \bibfield  {author} {\bibinfo {author} {\bibfnamefont {X.-D.}\ \bibnamefont
  {Ji}}\ and\ \bibinfo {author} {\bibfnamefont {R.~F.}\ \bibnamefont {Lebed}},\
  }\href {\doibase 10.1103/PhysRevD.63.076005} {\bibfield  {journal} {\bibinfo
  {journal} {Phys.\ Rev.}\ }\textbf {\bibinfo {volume} {D63}},\ \bibinfo
  {pages} {076005} (\bibinfo {year} {2001})},\ \Eprint
  {http://arxiv.org/abs/hep-ph/0012160} {arXiv:hep-ph/0012160 [hep-ph]}
  \BibitemShut {NoStop}%
\bibitem [{\citenamefont {Lebed}(2001)}]{Lebed:2001ic}%
  \BibitemOpen
  \bibfield  {author} {\bibinfo {author} {\bibfnamefont {R.~F.}\ \bibnamefont
  {Lebed}},\ }\href {\doibase 10.1103/PhysRevD.64.094012} {\bibfield  {journal}
  {\bibinfo  {journal} {Phys.\ Rev.}\ }\textbf {\bibinfo {volume} {D64}},\
  \bibinfo {pages} {094012} (\bibinfo {year} {2001})},\ \Eprint
  {http://arxiv.org/abs/hep-ph/0105218} {arXiv:hep-ph/0105218 [hep-ph]}
  \BibitemShut {NoStop}%
\bibitem [{\citenamefont {Dey}(2015)}]{Dey:2015rqa}%
  \BibitemOpen
  \bibfield  {author} {\bibinfo {author} {\bibfnamefont {B.}~\bibnamefont
  {Dey}},\ }\href {\doibase 10.1103/PhysRevD.92.033013} {\bibfield  {journal}
  {\bibinfo  {journal} {Phys.\ Rev.}\ }\textbf {\bibinfo {volume} {D92}},\
  \bibinfo {pages} {033013} (\bibinfo {year} {2015})},\ \Eprint
  {http://arxiv.org/abs/1505.02873} {arXiv:1505.02873 [hep-ex]} \BibitemShut
  {NoStop}%
\bibitem [{\citenamefont {K{\" o}rner}(2014)}]{Korner:2014bca}%
  \BibitemOpen
  \bibfield  {author} {\bibinfo {author} {\bibfnamefont {J.~G.}\ \bibnamefont
  {K{\" o}rner}},\ }in\ \href {\doibase 10.3204/DESY-PROC-2013-03/Koerner}
  {\emph {\bibinfo {booktitle} {{Proceedings, Helmholtz International Summer
  School on Physics of Heavy Quarks and Hadrons (HQ 2013): JINR, Dubna, Russia,
  July 15--28, 2013}}}}\ (\bibinfo {year} {2014})\ pp.\ \bibinfo {pages}
  {169--184},\ \Eprint {http://arxiv.org/abs/1402.2787} {arXiv:1402.2787
  [hep-ph]} \BibitemShut {NoStop}%
\bibitem [{\citenamefont {Aloni}\ \emph {et~al.}()\citenamefont {Aloni},
  \citenamefont {Grossman},\ and\ \citenamefont {Soffer}}]{Aloni:2018ipm}%
  \BibitemOpen
  \bibfield  {author} {\bibinfo {author} {\bibfnamefont {D.}~\bibnamefont
  {Aloni}}, \bibinfo {author} {\bibfnamefont {Y.}~\bibnamefont {Grossman}}, \
  and\ \bibinfo {author} {\bibfnamefont {A.}~\bibnamefont {Soffer}},\
  }\href@noop {} {\ }\Eprint {http://arxiv.org/abs/1806.04146}
  {arXiv:1806.04146 [hep-ph]} \BibitemShut {NoStop}%
\end{thebibliography}%

\end{document}